# High-order elastic multipoles as colloidal atoms

Bohdan Senyuk,[1] Jure Aplinc,[2] Miha Ravnik,[2,3] and Ivan I. Smalyukh[1,4,5,*]

[1]*Department of Physics and Soft Materials Research Center, University of Colorado, Boulder, Colorado 80309, USA*

[2]*Faculty of Mathematics and Physics, University of Ljubljana, Jadranska 19, 1000 Ljubljana, Slovenia*

[3]*J. Stefan Institute, Jamova 39, 1000 Ljubljana, Slovenia*

[4]*Department of Electrical, Computer and Energy Engineering, University of Colorado, Boulder, Colorado 80309, USA*

[5]*Renewable and Sustainable Energy Institute, University of Colorado, Boulder, Colorado 80309, USA*

*Email: ivan.smalyukh@colorado.edu

**Abstract**

Achieving and exceeding the diversity of colloidal analogs of chemical elements and molecules as building blocks of matter has been the central goal and challenge of colloidal science ever since Einstein introduced the colloidal atom paradigm. Recent significant advances have been achieved by exploiting the powerful machinery of DNA hybridization to assemble colloidal particles, but robust physical means of defining colloidal atoms and molecules remain limited. Here we introduce the physical design principles that allow for defining high leading-order elastic colloidal multipoles that emerge when particles with controlled shapes and surface boundary conditions are introduced into a nematic host fluid. Combination of experiments and numerical modeling of equilibrium field configurations, with a subsequent spherical harmonic expansion, allows us to systematically probe elastic multipole moments, bringing analogies with electromagnetism and structure of outmost occupied shells of atomic orbitals. We show that, at least in view of the symmetry of the "director wiggle" wave functions, the diversity of elastic colloidal atoms can far exceed that of known chemical elements.

## I. INTRODUCTION

Colloids are ubiquitous in everyday life and can be found everywhere from industrial, highly technological materials to commonly used health care and nutrition products [1-3]. Macroscopic structure and physical properties of self-assembled colloidal systems can be tuned by changing the interaction between their building blocks. Thus, designing new colloidal particles as atom-like building blocks can enable self-assembly of new complex artificial materials with desired properties, including the ones not encountered in nature. This "bottom-up" approach caused an explosion in the development of various kinds of colloidal particles [1-5]. Geometry of the colloidal particles is often used to define the directional interactions between them, similar as a valence of atoms determines the bonds that the atoms can form within molecules and crystals [6]. Directional bonding of colloidal particles has been pursued in many systems of colloidal "atoms" with varying shape and geometry [6], patchy features [7-9], and via various chemical functionalizations and DNA-hybridization [7, 10-14] of nano- and micro-particles. Elastic multipoles induced by colloidal particles dispersed in liquid crystals (LCs) provide yet another promising approach of defining controlled directional interactions of nematic colloidal "atoms" [15-31], with the design principles often building on the analogy of these multipoles with their electrostatic counterparts. When designing self-assembly of colloidal superstructures, the conditions of certain multipolar charge distributions, as well as the nature of interaction between them, can provide insights into how nematic colloids can be controlled [15]. For example, similar to electrostatic charge distributions, odd moments of elastic multipoles are expected to vanish when $\mathbf{n}(\mathbf{r})$ is symmetric about the particle center and a plane orthogonal to the far-field director, as in the cases of elastic quadrupoles [17-19] and hexadecapoles [31], whereas both odd and even moments should be present for particles with asymmetric $\mathbf{n}(\mathbf{r})$, such as elastic dipoles [16,18]. The design of elastic colloidal multipoles can potentially not only take advantage of such symmetry considerations, but also build on the versatile means of controlling $\mathbf{n}(\mathbf{r})$ by surfaces of colloidal inclusions with elaborate geometry and topology [15]. However, the general physical principles and feasibility of "on-demand" achieving a diverse variety of elastic colloidal multipoles remain unknown.

Colloidal particles locally distort the uniform alignment of the director field in a nematic LC host medium, prompting elasticity-mediated interactions between them, which tend to minimize the system's free energy [16]. Surface boundary conditions on the colloidal particles play an important role, often introducing bulk and surface line and point defects [15-31]. This gives rise to interactions between the colloidal particles, which tend to arrange such that the energetically costly distortions can be shared, and essentially resemble interactions of electrostatic multipoles [16, 18-21]. However, mainly only colloidal elastic dipoles [22, 23] and quadrupoles [17, 22, 24, 25] were studied, whereas the higher order multipoles were rarely considered, although recently the conic degenerate anchoring boundary conditions [30] enabled observation of hexadecapolar (16-pole) LC colloids [31]. Beyond the electrostatic analogy, elastic multipoles also share the mathematical description in terms of spherical harmonics with chemical elements (Fig. 1), whereby the elastic monopoles, dipoles, quadrupoles and octupoles have atomic analogs with the structure of filled *s*-, *p*- *d*- and *f*-orbitals [15,31]. Since none of the known chemical elements have filled orbitals higher than *f*, colloidal "atoms" in the form of elastic hexadecapoles and higher order multipoles have the potential to go beyond and, thus, could provide completely new insights and means of realization of new breeds of composite materials. Moreover, such artificial atoms and molecules could be used not only to define symmetry and structure of mesoscopic composite material systems, but also their physical properties. Indeed, the analysis of electric and magnetic multipoles [32] is also performed in optical metamaterials, as a central way to characterize the interaction of the electromagnetic fields with the material, which today underpins some of the most important technologies, ranging from telecommunications to data storage and light-assisted manufacturing [33, 34]. Within this approach, electromagnetic media are described as a set of point-like multipole sources, consisting of electric, magnetic and toroidal multipoles [33, 34]. Such metamaterials are typically nanofabricated, which limits their utility and calls for the development of means to self-assemble them from colloidal meta-atoms.

In this work, we uncover the physical mechanisms that may allow for an on-demand control of the leading-order elastic multipoles in LC colloids. We systematically demonstrate how shape and boundary conditions on colloidal particle surfaces determine the structure of elastic distortions, which allows us to identify the "design rules" for obtaining desired elastic multipoles and, thus, also the ensuing elasticity-mediated interactions. This may allow for developing pre-programmed composite metamaterials that self-assemble to yield the desired mesoscopic structure and physical properties.

## II. ELASTIC MULTIPOLES IN NEMATIC LIQUID CRYSTALS

Multipole expansion is an approach that represents a distinct complex spatially varying field – electric, magnetic, gravitational, material- as a series of elementary contributions – the multipoles [32, 33, 35]. Multipoles represent the magnitude and spatial profile of the basic sources of the fields, typically in some small region, to give the full fields in more distant regions. Multipole moments usually consist of inverse powers of the distance from the sources and angular dependent terms with the key assumption that only some – usually selected lowest order - multipoles are sufficient to adequately describe the full variability of the considered fields. The existence or non-existence of some multipoles not only has profound importance for the properties of the fields, but can even determine fundamental laws. One prime example is the nonexistence of magnetic monopoles, which fundamentally determines the structure of Maxwell's equations of electromagnetism, one of the fundamental laws of Nature [36]. Another prime example are atomic orbitals in atoms, where the quantum numbers of the orbitals reflect the corresponding underlying multipole-type nature of the atoms [37].

We explore elastic multipoles in the material orientational field of nematic complex fluids, with a central distinction that these multipoles can be directly measured and determined with optical and material science techniques and, moreover, that usually non-common leading-order multipoles such as 16-pole, 32-pole and even 64-pole can be realized (Fig. 1). The ordering field of nematic complex fields -into which multipoles will be imprinted by colloidal particles of designed surface-imposed ordering - is determined by an effective total free energy, which in full Landau-de Gennes form consists of effective nematic bulk elastic and ordering and surface anchoring terms [38] (see the Appendix A). The leading contribution that can transfer interaction, e.g. between multiple colloidal particles - colloidal atoms - in analogous way as electromagnetism in atomic orbitals, is the nematic elasticity which in the elementary one-elastic-constant form can be written as

$$f_E = \frac{1}{2} K \sum_{\mu=x,y} (\nabla n_\mu)^2, \tag{1}$$

where $f_E$ is elastic free energy density, $K$ is the single average Frank elastic constant and $n_\mu$ ($\mu = x, y$) are director components perpendicular to the far field direction ($z$ direction). This formulation of the free energy density relies on the crucial assumption of roughly uniform director field $\mathbf{n}(\mathbf{r}) \approx (n_x, n_y, 1)$, with small $n_x, n_y \ll 1$, which becomes justified at large distances from the localized source of nematic distortions, such as our colloidal particles. The elementary solutions of the nematic field around localized sources - the elastic multipoles - can be introduced by minimizing the free energy density (Eq. 1) with the Euler-Lagrange formalism giving Laplace equations:

$$\nabla^2 n_\mu = 0, \tag{2}$$

which in full 3D are separable and can be analytically solved as a series, i.e. as a summation over the elastic multipoles:

$$n_\mu(r,\theta,\varphi) = \sum_{l=0}^{\infty} \sum_{m=-l}^{+l} q_{lm}^\mu \frac{R_{\text{eff}}^{l+1}}{r^{l+1}} Y_l^m(\theta,\varphi), \tag{3}$$

where $\theta$ is polar and $\phi$ azimuthal angle, $Y_l^m(\theta,\phi)$ are spherical harmonics, $q_{lm}^\mu$ are dimension-free elastic spherical multipole coefficients, $l$ determines the order of a multipole as $2^l$-th pole, $-l \leq m \leq l$, and $R_{eff}$ is the characteristic scale of the multipole (given in our case by the effective size of the particle). Using orthogonality of spherical harmonics, the multipole moments $q_{lm}^\mu$ can be determined with the following integral

$$q_{lm}^\mu = \int_0^{2\pi} \int_0^\pi n_\mu(r,\theta,\phi) \frac{r^{l+1}}{R_{eff}^{l+1}} Y_l^{m*}(\theta,\phi) d\theta d\phi, \qquad (4)$$

which is particularly useful for calculating the multipoles numerically. Clearly, this integration must be performed at the radius that is large enough to satisfy the assumptions of $n_x, n_y \ll 1$ and $n_z \approx 1$. In practice the director field $n_\mu$, obtained from experiments or modelling, is defined only in discrete points ($\theta_i, \phi_j$); therefore, the calculation of the multipole coefficients reduces the discrete Fourier and Legendre transform on selected spherical grid (Appendix A). Figure 1 shows examples of elastic multipoles and relates the symmetry of associated director distortions to the analogous descriptions of charge distributions in electrostatics and electron wave functions in the description of electron shells of chemical elements.

### III. METHODS, TECHNIQUES AND MATERIAL SYSTEMS

#### A. Materials

We used a room temperature nematic LC 4-cyano-4'-pentylbiphenyl (5CB, from Frinton Laboratories, Inc.) or nematic mixture E7 (EM Industries) as a colloidal host medium. To define localized director distortions in the LC, mimicking symmetry of point sources of various multipoles (Fig. 1), we used a series of different colloidal particles. Gourd-shaped polystyrene dimer particles with two lobes of different diameter $2r_a \approx 2.5$ μm and $2r_b \approx 1.25$ μm (Figs. 2, 3) were synthesized using a modified seeded polymerization technique [39-41] and first dispersed in ethanol before introducing them into the nematic host. LC molecules aligned tangentially at the surface of the small lobe and exhibited a conic alignment at the surface of the large lobe. To have a tangential alignment of LC molecules at the particle's surface, we used superparamagnetic beads (SPMBs, Dynabead M450, Invitrogen) with a nominal diameter of ≈4.5 µm, which contained ferromagnetic nanoparticles embedded into a highly cross-linked epoxy [42, 43]. Alternatively, polystyrene spheres DC-05 (Thermo Fisher Scientific, Inc.) with a diameter of ≈5.3 µm also exhibited tangential boundary conditions. Glass particles (Thermo Fisher Scientific, Inc.) with a diameter of ≈5.1 µm treated with an aqueous solution (0.05 wt%) of N,N-dimethyl-N-octadecyl-3-aminopropyl-trimethoxysilyl chloride (DMOAP) exhibited perpendicular surface anchoring boundary conditions. All colloidal particles were dispersed in a LC host either via mechanical mixing or solvent exchange, producing dilute colloidal dispersions. After ~ 5 min sonication to break apart pre-existing aggregates, these colloidal dispersions in the LC state were filled in-between two glass plates spaced by glass spacers setting the gap thickness $d \approx$ 15-60 µm. Planar surface boundary conditions at confining substrates were set by unidirectionally rubbed thin films of spin-coated and crosslinked polyimide PI2555 (HD MicroSystem) [31]. To minimize spherical aberrations in experiments involving high numerical aperture (NA) immersion oil objectives, one of the used cell substrates was 0.15-0.17 mm thick, as needed for high-resolution imaging.

### B. Experimental techniques

An experimental setup assembled around an inverted Olympus IX81 microscope was used for optical bright-field and polarizing microscopy observations with a 100× (NA = 1.4) oil objective. To study distortions of the director field caused in a uniform nematic LC background by colloids, we also utilized a polarimetric imaging setup (Appendix C) integrated with the same optical microscope. Optical manipulations and assisted assembly of colloidal particles were realized with a holographic optical trapping system [44, 45] operating at a wavelength of $\lambda$ = 1064 nm and integrated with our optical microscope. Rotational manipulation of magnetically functionalized colloids was achieved using an in-house custom built holonomic magnetic manipulation system integrated within the same setup [43]. Translational and rotational motion of colloidal particles was recorded using a CCD camera (Flea, PointGrey) at a rate of 15 frames per second and the exact spatial positions and orientations of colloidal particles as a function of time were then determined from captured sequences of images using motion tracking plugins of the ImageJ (obtained from the National Institute of Health) analyzing software.

### C. Methods and procedures of numerical modelling

Elastic multipoles in our study were formed using dimers or assembly of spherical particles with the same or different anchoring and dimensions. In numerical modelling of gourd-shaped colloidal particles comprised of two dissimilar spheres (Fig. 2), the radius $r_a$ and position $d_a$ of the lower larger sphere were kept constant, whereby the radius $r_b$ and position $d_b$ of the upper smaller or equal sphere were varied to achieve different structures (see Appendix B2). The anchoring on the spheres was defined as strong and planar degenerate on the upper sphere and as conic degenerate on the lower sphere (Fig. 2). Both planar degenerate and conic degenerate anchoring impose distortions of the director field, which in the studied equilibrium structures are rotationally symmetric with respect to the $z$-axis.

The total free energy $F$ (Appendix A) was minimized numerically by using an explicit Euler relaxation finite difference scheme on a cubic mesh [46]. Material parameters of typical nematic LCs were used in the calculations: $L = 4 \times 10^{-11}$ N, $A = -0.172 \times 10^6$ Jm$^{-3}$, $B = -2.12 \times 10^6$ Jm$^{-3}$, $C = 1.73 \times 10^6$ Jm$^{-3}$ [46]. Simulations were performed on a square grid consisting of $400 \times 400 \times 400$ simulation points. For gourd-shaped dimers, composite colloids consist of two spheres, the bottom one has 100 points in diameter, whereas the upper sphere has from 20 to 100 points in diameter and is gradually moved upwards in steps of 10 points to mimic a broad range of shapes of dimer particles that can be obtained in experiments [39-41]. The composite colloids consisting of different number of equally sized spheres have 50 points in diameter. We assume fixed homeotropic anchoring on the cell surfaces and strong ($W = 10^{-2}$ Jm$^{-2}$) conic, homeotropic or planar degenerate anchoring on the composite colloids.

Multipole expansion was numerically performed with Gauss-Legendre algorithm, which was implemented via the numerical library `SHTns` [47, 48]. Several optimizations were used to achieve maximum efficiency [47], including the fast Fourier transform from the library `FFTW` [49, 50] to improve accuracy and speed. The main advantage of this library is the efficient on-the-fly computation of the Legendre-associated functions. Also, the algorithms implemented in `SHTns` are of high order accuracy $O(N^3)$, where $N$ is the number of calculation points [47].

## IV. RESULTS

### A. Elastic multipoles induced by composite colloids of dissimilar spheres

The aim of this work is to systematically investigate how high-order elastic multipoles can be induced by colloidal particles with varying shape and boundary conditions. Dimers of spheres are interesting sources of elastic distortions because they can be mass-synthesized using wet chemistry approaches [39-41] (making them relevant for composite material fabrication) while also allowing for more complex director distortions than what can be induced by individual colloidal spheres. We first study elastic multipoles formed around colloidal particles comprised of two dissimilar spheres (Figs. 3, 4) having different sizes, composition and anchoring. In the first case, gourd-shaped dimer colloidal particles consist of two lobes of different diameter and with different surface anchoring boundary conditions for the LC director (Fig. 3a-f). The dissimilar anchoring is defined through the particle synthesis, in which cross-linked polystyrene spherical seeds (a smaller lobe) were swollen with styrene and the elastic contraction of the cross-linked polystyrene expels styrene out of the swollen seeds to give rise to the second (larger) lobe [39-41]. As a result, the smaller particle's lobe has tangential anchoring and the larger lobe has a conic anchoring [30, 31]. Figure 3a-c shows POM textures of such particles in the LC, from which the director structure around the particle (Fig. 3d) was experimentally deduced. Gourd-shaped dimers align with their cylindrical symmetry axis parallel to the far-field director $\mathbf{n}_0$. These nematic colloids induce two surface point defects (called "boojums"), one at the south pole of a large lobe and another at the north pole of a smaller lobe, as well as a surface defect loop at the equator of the larger lobe (Fig. 3d,e). Additionally, a singular defect loop is visible spanning around the neck of a dimer particle, where two lobes come to the contact, which is due to the mismatch in the alignment of molecules at both lobes in the point of contact. Interestingly, the presence of this defect loop can be avoided by adjusting a distance $d_b$ (Figs. 2, 5) so that the particle geometry is fully compatible with the boundary conditions at the two lobes, as we will show below using numerical calculations. The induced configuration of the director is complex and the number of reversals of the director tilt with respect to the far-field director $\mathbf{n}_0$ at the surface of the particle is the same as in an elastic 64-pole (compare Fig. 3d,e and Fig. 1 for $l=6$). However, the structure also lacks symmetry with respect to the plane orthogonal to $\mathbf{n}_0$, which is not the case for the pure elastic 64-pole (see the analytical ansatz in Fig. 1). While the strengths of different elastic multipoles for such particles can be assessed numerically (see below), also the elastic interactions between such gourd-shaped dimers provide important insights (Fig. 3h-j). The highly anisotropic elastic interactions depend on the multipole magnitude and orientation (with respect to $\mathbf{n}_0$) of the separation vector connecting centers of two interacting gourd-like dimers. There are many narrow zones of attractive interactions separated by zones of repulsive interactions (Fig. 3g). For a colloidal particle with a 64-pole strongly dominating (or pure), the angular dependence would consist of 24 such zones, 12 attractive and 12 repulsive (Fig. 3g). However, the experimental angular diagram (Fig. 3h) is more complicated, pointing out a richer behavior that can be understood by presence of other non-zero elastic multipoles and consistent with the lack of symmetry plane orthogonal to $\mathbf{n}_0$. The elastic pair interaction potential between colloidal particles extracted from the experimental distance versus time dependencies is several hundreds of $k_BT$, with the interaction force approaching ~1 pN near their full contact. Fitting the experimental interaction potentials (Fig. 3i,j) with an expression for the colloidal pair-interaction energy within the multipolar approach [31] $U_{\text{int}} = \frac{4\pi K}{k_B T} \sum_{l,l'} b_l b_{l'} (-1)^{l'} (l+l')! \frac{R_{\text{eff}}^{l+l'+2}}{r^{l+l'+1}} P_{l+l'}(\cos\theta)$, where $P_{l+l'}(\cos\theta)$ are the Legendre polynomials, yields the coefficients corresponding to the strength of elastic multipole moments (Fig. 3). We find that the hexadecapolar moment $b_4$ is still strongly pronounced, which is due to the larger lobe with conical anchoring (Fig. 6). Although different approaches can be used for fitting (Fig. 3), it is clear that the role of 64-pole ($b_6$) becomes significant at small inter-particle distances $r$. The interaction potential of gourd-shaped particles (Fig. 3i,j) cannot be fit using expressions with only quadrupolar ($b_2$) or hexadecapolar ($b_4$) nonzero coefficients. The angular dependence of elastic interactions of two gourd-shaped particles - as expressed by different relative magnitudes of thee multipole coefficients - is very sensitive to the geometrical parameters of a dimer, like the relative dimensions and overlap of the dimer lobes, which is calling for the need of establishing design principles for expressing the desired leading-order multipoles, as we do numerically below.

Another model colloidal object in our study is a dimer particle consisting of two spheres with dissimilar size and anchoring boundary conditions, which is made from SPMB and glass spheres (Fig. 4). SPMBs are somewhat smaller as compared to glass particles (Fig. 4). Using the laser tweezer, two different particles were placed close to each other and the LC around them was locally melted to the isotropic state by converting the infra-red laser irradiation to heat through absorption, irreversibly forming a colloidal dimer bound by van der Waals forces. The LC was then quenched back to the nematic state. The resulting dimer aligns with the symmetry axis along $\mathbf{n}_0$ (Fig. 4). SPMBs look somewhat brownish because of the ferromagnetic nanoparticles in the resin within them and show the planar anchoring for LC director. On the contrary, the treated glass particles impose perpendicular alignment of LC molecules at their surfaces. The resulting director structure contains a surface point defect on the pole of a SPMB sphere and a bulk disclination loop, called "Saturn ring", at the equator of a glass sphere (Fig. 4e-g). Analysis of experimental and numerical **n(r)**, including the corresponding color-coded diagram of the $n_x$ sign alternation at the surface of the dimer, reveals the structural similarity with an elastic octupole with $l=3$ (compare Fig. 4e and Fig. 1). Other structures of **n(r)** around such dimers were also observed (Fig. 4d,h), within which the accompanying disclination loop shifted from the equator of the glass particle and resided near the SPMB sphere, just above the contact point of two particles, effectively generating a dipolar director structure. This diversity of multipolar structures [7] that can be induced by colloidal dimers calls for identifying parameters that can lead to on-demand formation of various multipoles.

Figure 5a shows numerically calculated structures of **n(r)** induced by colloidal particles composed of two dissimilar spherical colloidal objects. The radius of the upper sphere $r_b$ is varied from 0 to the size of the lower sphere $r_a$ in steps of $r_a/5$. The position of the upper sphere is also varied and is gradually immersed into the lower sphere in steps of $r_a/5$, where, in the limiting regimes, the upper sphere is completely contained (left side of the pyramids) or barely touches the lower one (right side of the pyramids). Conic anchoring on the lower sphere introduces additional parameter $\alpha$ (Fig. 2), that corresponds to the tilt angle of the conical anchoring. In the limiting case when the conic anchoring angle $\alpha$ reaches 90° (Fig. 5a), the director is locally oriented perpendicular to the surface of a lower spherical lobe. This homeotropic anchoring induces a disclination loop defect "Saturn ring" on the lower sphere. The tangential anchoring on the upper sphere induces a boojum at its north pole. Another loop of reduced degree of order can emerge in the neck, where the two spheres with distinct anchoring meet. It is induced by a mismatch between the two easy-axis directions imposed at the contact of the two surfaces by two different boundary conditions, measured with angle $\beta$ (Fig. 2). The mismatch angle $\beta$ characterizes the effective surface-imposed frustration in the defect region within the neck and depends on both the conic anchoring angle $\alpha$ as well as on the angle at which the two spheres intersect. The neck region with the reduced degree of order vanishes when geometry of the composite colloid (size and displacement of the upper sphere) assures intersection of the two spheres at 90° setting mismatch angle $\beta = 0°$ (third diagonal from the right side of the pyramid in Fig. 5a). As also observed in experiments (Fig. 4d,h), the Saturn ring can change its vertical position and slide from equator of the lower colloid towards the neck, or even on the neck joining with the regular neck defect. This rearrangement occurs in the lower left region of the parameter-space pyramid (Fig. 5a), where the director field is found to be dipolar-like.

Figures 5b, 5c and 5d show the composite colloids and their corresponding field structures when conic anchoring boundary conditions respectively at angles $\alpha = 60°$, $\alpha = 40°$ and $\alpha = 20°$ are applied on the lower sphere. For these conic angles, the distinct nature of conic anchoring (Fig. 2) induces a ring defect at the equator of the lower sphere. The tangential anchoring on the upper sphere establishes a boojum defect at the surface of the colloidal particle. At the neck, where the two spheres intersect and the two regions with dissimilar anchoring meet, again the singular region of reduced order emerges. Its actual profile depends on the mismatch angle $\beta$, which is now conditioned by the conic anchoring angle $\alpha$ and composite colloid parameters $r_b$ and $d_b$. The neck region of reduced order completely vanishes when the two spheres intersect at an angle equal to the conic anchoring angle $\alpha$. Indeed, one can see that the neck region of reduced order effectively disappears for the composite colloids at third diagonal from the left side of the pyramid in cases $\alpha = 60°$ (Fig. 5b), $\alpha = 40°$ (Fig. 5c) and for second diagonal from the left in case of $\alpha = 20°$ (Fig. 5d). When the conic anchoring angle $\alpha$ is small (Fig. 5d), the anchoring on the lower colloid is nearly tangential. Consequently the effective Saturn ring distortions become less pronounced, whereas the distortion of the neck region of reduced order becomes stronger at large separations $d_b$ (right edge of the pyramid on Fig. 5d). We have analyzed elastic multipole moments of experimental (Figs. 3, 4) and simulated (Fig. 5) structures of composite particles using decomposition on spherical multipoles with the `SHTns` numerical library (Appendix A), as detailed below.

Figure 6 shows the spherical multipole coefficients of all simulated composite colloids at various angles $\alpha$. The dipolar coefficient $q_{11}^x$ is generally weak and positive for low $\alpha$; however, a prominent peak is observed at $\alpha = 90°$ for the structures close to $r_b / r_a = 4/5$, $d_b / r_a = -2/5$ (see Fig. 5a). Consistent with experiments, it is energetically favorable that the Saturn ring defect moves from the equator of the lower particle to the neck, effectively joining with the neck defect. Defect in the neck only weakly distorts the surrounding director field (Fig. 5a); therefore, the defect playing a key role is the boojum at the north pole of the upper particle, resulting in a dipolar director deformation. The dipolar multipole is still present in this region, even at lower conic anchoring angles $\alpha$. The quadrupolar coefficient $q_{21}^x$ is strongest and negative for homeotropic bottom particles, where the Saturn ring is the sole defect, and becomes weaker upon decreasing the conic anchoring angle. At $\alpha = 40°$ negative contribution vanishes and strong positive quadrupole moment emerges for spherical colloids with nearly tangential anchoring at $\alpha = 20°$, where two boojums emerge at both poles. The transition between a positive and negative quadrupole moments can be understood by comparison with Fig. 1. An octupole moment $q_{31}^x$ is always positive and strongest for large upper spheres with conic anchoring angles near $\alpha \sim 90°$. This composite colloid has Saturn ring on the lower colloids equator, and a strong boojum on the upper particle's north pole (Figs. 4 and 5a), whereas the role of neck defect in defining multipoles often can be negligible. The hexadecapole moment $q_{41}^x$ is positive and has similar strength and pattern at all angles $\alpha$. The hexadecapolar contribution of the homeotropic spherical particle emerges because the lower particle is off-centered within the simulation cell, which affects the high multipoles. The hexadecapolar coefficient is present also at angles other than $\alpha = 90°$. The best candidates for "pure" hexadecapolar elastic colloids are spheres with conic anchoring, which preserves the Saturn ring, but also induce boojums at the poles, which is in good agreement with the recent experiments [31]. The high-order 32-polar coefficient $q_{51}^x$ is generally weak. It is the strongest for composites with larger upper sphere and high anchoring angle, which produces Saturn ring on the lower particle and boojum at the north pole. The 64-polar coefficient $q_{61}^x$ appears to be most prominent for the spherical particles with homeotropic boundary conditions. As in case of the hexadecapole, it is related to the strong dipole and quadrupole. Besides the simple homeotropic sphere, the 64-polar contribution is maximized for composites with large upper sphere and moderate anchoring angle $\alpha$, which provides boojum at the upper particle, neck defect, Saturn ring at the equator of the lower particle and boojum at the lower particle (Fig. 3).

By tuning the geometry and anchoring on the composite particle with two dissimilar spheres one can maximize certain multipole moments and suppress the others. For example, homeotropic particle with dimensions near $r_b / r_a = 4/5$, $d_b / r_a = -2/5$ acts as a strong dipole. Expectedly, colloidal spheres with homeotropic or nearly tangential anchoring exhibit a strong quadrupolar moment. An octupole moment is maximized for composite particles, which consists of a large upper sphere with tangential anchoring touching the lower sphere with homeotropic boundary conditions (Fig. 4). A hexadecapole moment is the strongest for single spherical particles with conic anchoring at $\alpha = 40°$.

The comprehensive multipolar analysis presented in Fig. 6 shows that different leading-order multipoles can be created simply by engineering distributions of director field, closely mimicking the corresponding analytical ansatzes (Fig. 1). Clearly, to have a desired leading order $2^l$-pole with $m=1$ and structure axially symmetric with respect to $\mathbf{n}_0$, the director tilt (and the sign of $n_x$) must change $2l$-times when one circumnavigates around the colloidal object once (Fig. 1), with additionally the structure being symmetric with respect to a plane orthogonal to $\mathbf{n}_0$ for all even-order multipoles. With color presentations of director tilt (Fig. 1), it suffices to count the yellow-blue changes of colors to assure that this is the case. Our findings also reveal how higher-order multipole can be expressed by introducing additional defects, either boojums or disclination rings. For the case of dissimilar dimer particles, our analysis shows that a structure with one boojum tends to exhibit a leading-order dipole moment ($l=1$, $2^l=2$), whereas presence of a Saturn ring tends to yield a quadrupole ($l=2$, $2^l=4$) (see Fig. 1). The octupole ($2^l=8$) is produced by combining the boojum ($2^l=2$) and the Saturn ring ($2^l=4$) (Fig. 4): $2 \times 4 = 8$, while hexadecapole ($2^l=16$) can be thought of as a superposition of two quadrupoles ($2^l=4$): $4 \times 4 = 16$ or comprising a quadrupole ($2^l=4$) and two boojums ($2^l=2$): $2 \times 4 \times 2 = 16$. These examples yield one recipe to generate an arbitrary high-order multipole $M$ ($M$-pole) by the dissimilar dimer particles:

$$M = 2^i \times 4^j, \tag{5}$$

where $i$ is the number of boojums and $j$ is the number of Saturn ring defects. Here we consider defects solely as building blocks of director structures at which the vectorized $x$-component of the director changes its sign (Fig. 1). Also, we disregard the fact that, in general, beside boojums and Saturn rings, point defects (e.g. such as hyperbolic hedgehog) can form in nematic colloids, for which one would expect that they break symmetry in multipoles in analogous way as boojums. Finally, the detailed numerical exploration of strengths of multipole moments (Fig. 6) and the experimental example shown in Fig. 3 also show that for desired multipoles to dominate also geometric parameters and boundary conditions need to be optimized to reduce the other competing multipole moments. The symbols like circles and diamonds in Fig. 6 mark regions in the parameter space of colloidal dimers where such pure or dominant leading-order multipoles can be achieved.

### B. Elastic multipoles induced by composite colloids of similar spheres

The dimer particles studied above are examples from a large family of colloidal objects that can be used to induce desired elastic multipoles. As another experimentally accessible example, we study composite colloidal inclusions consisting of spherical constituents with similar size and boundary conditions. We use epoxy-based spherical SPMBs to fabricate dimers, trimers, tetramers and so on, all consisting of similar spherical particles, in order to demonstrate how they induce director structures corresponding to various higher-order elastic multipoles. In bright-field microscopy textures, epoxy-based SPMBs look brownish due to the light absorption by magnetic nanoparticles embedded within them. SPMBs define planar boundary conditions for LC molecules as revealed by the polarizing optical microscopy textures (Fig. 7a,b). Embedded magnetic nanoparticles allow for a facile control of SPMBs with a magnetic field. To form composite linear particles consisting of two or more SPMBs, individual SPMBs were placed nearby each other using laser tweezers and the LC was locally melted, again with tweezers, to the isotropic state. While in the melted isotropic area, particles where arranged into the linear chains of two or more touching particles using laser tweezers, and then chains were aligned parallel to the cell's rubbing direction using holonomic magnetic control (Fig. 7c,d). They irreversibly formed dimers, trimers or tetramers bound together by van der Waals forces. After the locally melted LC was quenched from isotropic back to the nematic state, the magnetic field holding the particle chains was removed, leaving the chains stable and oriented along $\mathbf{n}_0$. The alternation of director tilt at particle surfaces with respect to $\mathbf{n}_0$ revealed by different modes of microscopy is consistent with that in the ansatzes of sources of high-order elastic multipoles (compare Fig. 7h,l and Fig. 1). We also employed the polarimetric imaging (Appendix C) of director distortions around colloidal particles (Fig. 7m-o), results of which were consistent with polarizing micrographs and numerically calculated director structures. While a single SPMB induces an elastic quadrupole, a pair of SPMBs in a chain oriented along $\mathbf{n}_0$ has a strong elastic hexadecapole moment (compare Fig. 7h and Fig. 1 for $l=4$), which can be the strongest leading-order elastic multipole for certain parameters. Similarly, a chain of three particles can have a strongly pronounced 64-pole (compare Fig. 7l and Fig. 1 for $l=6$).

High-order multipoles can be also achieved by chains of irreversibly bound spherical particles with conic or homeotropic anchoring on their surfaces. We use numerical calculations to probe the geometric and boundary conditions parameter space for such particles (Fig. 8) and the ensuing strengths of elastic multipole moments. Defect configurations and the director field structures depend on the conic anchoring angle. Within the composite particles, each homeotropic sphere generates Saturn ring defect around its equator (Fig. 8a). However, when the conic anchoring angle $\alpha$ is diminished, defects arise at the neck of the particle and at each free pole (Fig. 8a). With increasing $\alpha$ further, obtaining tangential anchoring,

Saturn ring vanishes and only neck defects and boojums remain. In the language of elastic multipoles this means that the single spherical particle transitions from negative quadrupole (i.e. with negative multipole coefficient $q_{21}^x$) to $2 \times 4 \times 2 = 16$ pole and then to the positive quadrupole [31]. A chain-like particle comprising two spheres transitions from a hexadecapole with a negative multipole moment ($4 \times 4 = 16$) for homeotropic anchoring to $2 \times 4 \times 4 \times 2 = 64$-pole and then to a hexadecapole with a positive multipole moment ($2 \times 4 \times 2 = 16$) for tangential anchoring. Chain particles from three spheres could induce strong -64-pole, 256-pole and +64-pole, while the particle comprised of four spheres can effectively act as -256-pole, 1024-pole and +256-pole, depending on boundary conditions. To summarize, with parameters for the desired leading-order multipoles optimized, colloidal oligomer chain particles with $N$ similar spheres can act as $-4^N$-pole in the case of homeotropic anchoring, $2 \times 4^N \times 2$-pole for moderate conic anchoring and as $2 \times 4^{N-1} \times 2$-pole when anchoring is tangential. We note that neck defects at moderate $\alpha$ are inconsequential for defining the strongest leading-order multipoles, because they cause only weak or no deformations, though they can affect the strength of higher-order multipoles.

Graphs in Fig. 8b show spherical multipole coefficients $q_{l1}^x$ for chain composite colloids with various conic anchoring angle $\alpha$. These coefficients were calculated such that the center of the interpolation sphere coincides with geometrical center of the symmetric chain composite colloids. Generally, particles show no or weak dipolar moment, except for homeotropic particle with four spheres, for which Saturn rings move upwards towards the necks and pole (Fig. 8a). Multipole moments with odd $l$ tend to be zero due to particle (and consequently director field) symmetry (Fig. 8b). Particles with only one sphere tend to show the strongest quadrupole moment $q_{21}^x$. A sphere with homeotropic anchoring forms a elastic quadrupole with a negative magnitude, whereas the sphere with tangential anchoring is an elastic quadrupole with a positive magnitude (Fig. 8b). Interestingly chain colloids of arbitrary number of spheres with conic anchoring angle $\alpha = 40°$ show no quadrupole moment. Colloidal particles with one sphere exhibit a strong positive hexadecapole moment $q_{41}^x$, which extends to multiple sphere particles, with conic anchoring angle below $\alpha = 40°$. When the anchoring angle exceeds conic anchoring angle $\alpha = 40°$, particles with multiple spheres transition into negative haxadecapole moments. Chain composite particles can have both positive and negative 64-pole moment, whereas structures in between have none. It appears to be the strongest for spherical particles with large conic anchoring angle $\alpha$ and composites of two or three spheres with low conic anchoring angle $\alpha$, as in the composite particles observed in the experiment (Fig. 7). Both 256- and 1024-pole ($q_{81}^x$ and $q_{10\,1}^x$) appear to be small, though reaching both positive and negative values as the geometric and surface anchoring parameters are varied. An interesting observation is that a sphere with homeotropic or tangential anchoring can act as strong elastic quadrupoles with opposite signs of moments. Conversely, spheres with conic anchoring at $\alpha = 40°$ exhibit a pure hexadecapole due to the superposition of opposite-charged quadrupole moments. 64-polar contribution is not significantly strong in general, however it is a prominently expressed multipole for a particle consisting of two spheres with anchoring angle $\alpha = 40°$.

Presented results for composite chain particles with up to two spheres are in a good agreement with predictions posed on the basis of arbitrary multipole creation principle discussed above (see Eq. 5). However the colloidal composites comprising three and especially four spheres are more complex. The apparent mismatch arises because the particles are long with respect to the typical range of deformation in the nematic LC, objects extended along the far-field director. Consequently, the corresponding distortions in the director field do not resemble the pure analytic multipoles (Fig. 1), but rather decompose into large number of multipoles. The problem could be mitigated by using oblate spheroids instead of spheres as building blocks of the chain colloids, putting the induced topological defects closer together, so that the alternation of director tilt could more closely mimic the corresponding ansatzes (Fig. 1).

## V. DISCUSSION

Our experimental results and numerical calculations show that one can design colloidal particles with different elastic multipoles, including the higher-order ones, by changing the shape and boundary conditions of constituent particles. For example, using dimer particles with different size and surface anchoring of constituent lobes (Fig. 3), we could design colloidal particles with enhanced 64-pole (a green circle in Fig. 6). Numerically calculated director fields (Fig. 5) and diagram (Fig. 6) for strength of multipole moments of such dimer particles provide insights for designing colloids with enhanced desired multipoles, showing how strongly pronounced multipoles of different order can be pre-selected by varying boundary conditions and particle geometry. Following a similar strategy, we also designed a colloidal dimer with octupolar-like configuration of the director field (Fig. 4) and the enhanced elastic octupolar moment (a blue circle in Fig. 6). It is interesting that the same arrangement of two constituent particles can also allow for defining a director field with enhanced dipolar moment (a red circle in Fig. 6) under different conditions (Fig. 4d,h). The calculated multipolar moment diagram includes also configurations with dominant or "pure" elastic multipoles as quadrupolar, octupolar and hexadecapolar (with the latter marked by a blue diamond in Fig. 6). As one can see from a diagram, the hexadecapolar elastic moment is pronounced to a smaller or larger extent in all configurations, and that the conditions for it to be a leading-order multipole can be created in multiple ways.

Composite particles with different higher order elastic multipoles can be also designed using colloidal oligomers formed by similar connected spheres (Fig. 7), which is consistent with the results of numerical calculations (Fig. 8). For example, a dimer of two particles with tangential surface anchoring, which each separately have a leading quadrupolar elastic moment, together give a rise to strongly enhanced hexadecapolar moment (marked by a red circle in Fig. 8). Three particles show the configuration of the director field with the director tilt reversals characteristic for a 64-pole with detectable corresponding elastic moment (Fig. 1).

The presented strategies allow for the design of colloids with a variety of elastic multipoles, including high leading-order multipoles like hexadecapoles and also "mixed" multipoles with strongly pronounced multipoles of different order. While having "pure" elastic multipoles is fundamentally interesting and has the advantage that they can be used for designing colloidal self-assemblies on the basis of corresponding well known interaction potentials, nematic colloids with "mixed" multipoles can cover even larger diversity of anisotropic elastic interactions (Fig. 3). The strategies described in this work, which involve dimers, trimers and oligomers of similar or dis-similar colloidal spheres, are just examples showing how the uniform alignment of the nematic host can be locally perturbed to mimic the corresponding ansatzes of elastic multipoles (Fig. 1). However, similar multipolar director distortions can be also achieved using colloidal objects with complex shapes obtained by means of photolithography [18,21,51] and two-photon-polymerization [52,53]. On the other hand, the concept of controlling surface anchoring on spherical constituents of composite colloidal objects that we present here can be extended to patchy particles [7-9,54], where different patches can exhibit different boundary conditions, and particles with controlled surface topography [43], surface charging [55] and chemical functionalization [15]. In nematic hosts, these highly tunable multipolar elastic interactions can be further enriched by weakly screened electrostatic monopole-like [56] and magnetic dipolar [57,58] interactions in cases of charged or magnetic particles. The ability of describing elastic, electrostatic and magnetic interactions as multipoles of different nature and order is a useful platform for designing LC colloidal composites. Interestingly, in this respect the electrostatic monopoles in LCs have been studied [56] but designing higher order electrostatic multipoles appears to be challenging so far. Differently, magnetic monopoles are considered impossible while dipoles can be easily obtained by using magnetically monodomain particles [57,58]. Elastic dipoles, quadrupoles, and hexadecapoles have been studied previously [15] and now the spectrum of accessible elastic multipoles is significantly broadened by this work. We envisage that this "zoo" of multipoles of different nature and order, which now by far exceeds the diversity of chemical elements similarly described by use of spherical harmonics [15], will be useful in "on demand" designing and realizing composite materials with desired structure and composition. In the case when nematic colloids have mixed multipoles with comparable strengths, although lower-order multipoles within them will define the behavior at large distance due to the inverse power type of scaling of elastic potential, the higher order multipoles can still significantly influence the behavior of elastic colloids at short center-to-center distances, which is where the details of self-assembled colloidal superstructures are defined.

Methodologically, our work is based on a combination of experiments and numerical modelling, where experimentally wet chemistry is used to variably create gourd shaped particles and a combination of optical microscopy techniques is used to determine their multipolar properties, whereas numerical modelling is based on phenomenological free energy minimization approach to calculate the ordering fields, and is then complemented by multipolar expansion algorithm into spherical harmonics. This approach can be effectively extended to other potential strategies of designing elastic multipoles discussed above, as well as supplemented by adding magnetic and electrostatic interactions [56-58]. Since the colloidal objects can have different compositions, including constituents made of noble metals [57], magnetic materials [42,43,57,58], semiconductor nanoparticles [55,56] and dielectric objects [17,18,21,24,31,51-53] (with means of defining boundary conditions for director on such colloidal objects already demonstrates [17,18,21,24,31,42,43,51-59]), we envisage that properties of the ensuing colloidal composite metamaterials can be pre-engineered by expanding the above described design toolkit to account for collective behavior of such assemblies enriched by plasmonic resonances [32-34], plasmon-exciton interactions [60,61], etc.

## VI. CONCLUSIONS

This work demonstrates realization of colloidal atoms from high-order multipoles based on geometrical and topological design of distortion fields in nematic colloidal fluids. The high-order multipolar colloidal objects are realized from elastic multipoles in the orientational director fields of nematic fluid that also can transfer inter-particle interactions of multipolar symmetry. We show realization of colloids with dipolar, quadrupolar, octupolar, hexadecapolar, 32-polar and even 64-polar multipole components, that we show not only can be controllably varied, but also designed by controlling particle shape and surface anchoring boundary conditions. Interestingly, we are also able to identify regimes –i.e. colloids with distinct geometrical and surface parameters- of 'pure' or leading-order multipoles, where a single multipole dominates and leads the structure, and 'non-pure' or 'mixed' multipoles, where various combinations of different multipoles are present on a single particle and determine the system.

More generally, this work is a contribution towards developing a novel -colloidal- matter that rather uniquely can go beyond the interaction types that are possible in the set of known atoms as determined by their orbitals. We show design of high-order multipoles, such as hexadecapole, 32-pole and even 64-pole, that can be mapped to atomic orbitals (subshells) of $l = 4$, 5 and 6, respectively, which do not have direct analogs in atomic and molecular systems. At this stage, our work is primarily centered around demonstrating the capabilities to realize individual particles – colloidal atoms, and basic interactions. In this work, the interaction range of high-order multipoles is also limited by relatively short range decay of high-order multipoles, but we believe that it is exactly by combining geometrical and topological approaches, that one could possibly open a field to beyond-atomic matter with novel material properties.

Our work shows that, similar to how we often think about high-order multipolar charge distributions in electrostatics, where high-order multipoles emerge from superposition of the lower-order ones when lower-order multipoles mutually cancel, high-order elastic multipoles can be designed by superimposing the lower-order ones and tuning conditions for cancelation of multipoles up to the desired leading-order one. The illustrative examples of this are composite colloidal particles, where each of them individually would induce a lower-order multipole, dimers, trimers, tetramers and oligomers of such particles can prompt creation of high-order multipoles under proper conditions (e.g. a dimer of colloidal quadrupoles can be arranged so that all lower-order multipoles but hexadecapolar cancel, making an elastic hexadecapole). These insights offer simple but powerful means for designing self-assembled colloidal composites.

**ACKNOWLEDGMENTS**

We acknowledge technical assistance of Owen Puls and discussions with B. Fleury, M. Tasinkevych, N. Wu and S. Copar. Research at CU-Boulder (B.S. and I.I.S.) was supported by the U.S. Department of Energy, Office of Basic Energy Sciences, Division of Materials Sciences and Engineering, under the grant DE-SC0019293 and by the National Science Foundation Grant DMR-1420736 (imaging and fabrication facilities). Research at FMF UL and IJS (J.A. and M.R.) was supported by Slovenian Research Agency Grants (Grant Nos. J1-7300, L1-8135, and P1-0099) and US Air Force Office of Scientific Research, European Office of Aerospace Research and Development (Grant No. FA9550-15-1-0418, and Contract No. 15IOE028).

**APPENDIX A: ELASTIC MULTIPOLES IN NEMATIC LIQUID CRYSTALS**

**1. Free energy**

An important method for studies of nematic structures is the Landau-de Gennes (LdG) free energy approach [38]. It is based on the full tensorial order parameter field $Q_{ij}$, which incorporates the orientation of the director **n**, orientation of the possible biaxial ordering relative to the director, scalar degree of order $S$ and biaxiality $P$. LdG modelling is a phenomenological approach which uses a tensor order parameter to construct a free energy functional *F*, which is also able to fully characterize the defect regions. We use one elastic constant approximation for the LdG free energy, which reads

$$F = \int_{LC} \left\{ \frac{A}{2} Q_{ij} Q_{ji} + \frac{B}{3} Q_{ij} Q_{jk} Q_{ki} + \frac{C}{4} (Q_{ij} Q_{ji})^2 \right\} dV$$
$$+ \int_{LC} \left\{ \frac{L}{2} \frac{\partial Q_{ij}}{\partial x_k} \frac{\partial Q_{ij}}{\partial x_k} \right\} dV \tag{A1}$$
$$+ \int_{Surf} \left\{ \frac{W}{2} (Q_{ij} - Q_{ij}^{(0)})(Q_{ij} - Q_{ij}^{(0)}) \right\} dS,$$

where LC denotes the integration over the bulk of the liquid crystal and Surf over the surface of the colloidal particles. The first term accounts for the variation of the nematic degree of order; *A*, *B* and *C* are material parameters. The second term penalizes elastic distortions in the nematic state, where *L* is the elastic constant. The final term in *F* is surface free energy, which accounts for the LC interaction with the surface of the colloidal particle, where *W* is the anchoring strength and we assume anchoring along preferred direction imposed by the leading eigen-pair of $Q_{ij}^{(0)}$ (i.e. with largest eigenvalue) [62]. The preferred direction of anchoring is set according to the type of the anchoring; note that the surface free energy in Eq. A1 imposes *uniform* anchoring along some distinct direction bot not degenerate (such as degenerate planar or conical). Nevertheless, in the work shown, the experimentally realized multipolar particles always exhibited rotational symmetry about the far field undistorted director, which makes the use of such uniform surface free energy appropriate and sufficient.

## 2. Nematic elastic multipoles

Nematic elastic multipoles are commonly known today and used in the literature as approximations for elastic distortion profiles of nematic orientational fields, the director, that surround colloidal particles. Nematic orientational fields can be calculated analytically only for selected, typically rather simple systems; however, in most cases the general solution cannot be obtained. In colloids, the key problem usually arises in the proximity of particle surfaces, where strong spatial gradients emerge in the nematic orientational fields, which is though different to typically small gradients away from particles. Therefore, to obtain analytical insight into nematic fields, the full Euler-Lagrange equations were simplified under selected assumptions (linearized) to be analytically solvable, eventually in the far-field in terms of elastic multipoles [22, 26, 28, 63].

The expansion to nematic elastic multipoles relies on the crucial assumption of roughly uniform director field $\mathbf{n}(r) \approx (n_x, n_y, 1)$, with small $n_x, n_y \ll 1$, where note that by definition, $\mathbf{n}$, is to be a unit vector field. Typically, such assumption can be justified at sufficiently large distances from colloidal particles (such as order of magnitude one particle radius away from the particle surface or can be even less). An additional assumption is also that the nematic elastic modes (i.e. elastic free energy) are described with one single elastic constant. Taking such approximations, the full nematic elastic free energy (second line in Eq. A1) can be simplified to the harmonic free energy $f_E = \frac{1}{2} K \sum_{\mu=x,y} (\nabla n_\mu)^2$, where we use notation $n_\mu$ ($\mu = x, y$) for components perpendicular to far field direction. The corresponding Euler-Lagrange equations are Laplace equations: $\nabla^2 n_\mu = 0$. The solution of Laplace equations is now sought in terms of series of multipoles.

The elastic multipoles can be introduced via spherical multipole moments or by using Green function [22], where the two approaches can be directly mapped one into another. For our work and analysis of results, it is convenient to use spherical harmonics as they can be readily determined by an expansion of the nematic director on a sphere that encloses the considered multipolar colloidal particles. Laplace equations for nematic director components $n_\mu$ ($\mu = x, y$) are separable in spherical coordinates and can be analytically solved, with their general solutions written as a sum of multipolar contributions

$$n_\mu(r,\theta,\varphi) = \sum_{l=0}^{\infty} \sum_{m=-l}^{+l} q_{lm}^\mu \frac{R_{\text{eff}}^{l+1}}{r^{l+1}} Y_l^m(\theta,\varphi),$$ where $\theta$ is polar and $\varphi$ azimuth angle, $q_{lm}^\mu$ are spherical multipole moments coefficients and $Y_l^m(\theta,\phi)$ are spherical harmonics. In order to extract distinct coefficient of selected multipole moment the orthogonality of spherical harmonics is used $\int_0^{2\pi} \int_0^{\pi} Y_l^m(\theta,\phi) Y_j^k(\theta,\phi) \sin\theta d\theta d\phi = \delta_{lj} \delta_{mk}$, where $\delta_{ij}$ is the Kronecker symbol. Note that the radius $r$ of the sphere at which the expansion is performed can be easily taken large enough to satisfy the assumptions $n_x$, $n_y \ll 1$ and $n_z \approx 1$.

In homogeneous background field (we use director along $z$ direction $\mathbf{n}_0 = \{0, 0, 1\}$), the symmetry of elastic multipolar distortions generated by particles is determined by the symmetry of the colloidal particles and their anchoring. In this work we have considered only particles invariant with respect to rotations about $z$ axis, which have no azimuthal contribution to $\mathbf{n}$, hence the director field imposed by a selected particle should be invariant with respect to rotations about the far-field director $\mathbf{n}_0$. This constraint sets monopole coefficient $A^\mu = 0$ and furthermore, implies that $q_{lm}^\mu$, with $m = \pm 1$ are the only non-vanishing coefficients in the expansion (Eq. 4) of the Cartesian director field components $n_\mu$ ($\mu = x, y$).

The Laplace equation has no inherent length scale; therefore, also no inherent length scale is present in the multipolar expansion as solution of the Laplace equation for the nematic director components. Nevertheless, clearly, already from the perspective of the dimensional analysis, multipoles have units of powers of a certain length scale. In our case of nematic elastic multipoles, we introduce this certain length scale as $R_{\text{eff}}$ (e.g. as introduced in Eq. 4), thus making the elastic multipolar coefficients $q_{lm}^\mu$ dimensionless to allow for comparison of the magnitudes of different multipoles. In colloidal systems, this certain length scale is naturally related to the particle size, which for spherical particles is the radius. However, for more complex shaped particles - like our gourd particles - it is less clear how to select this scale $R_{\text{eff}}$, especially if wanting to effectively compare multipole coefficients of particles of somewhat different shapes. Therefore, some selection has to be made according to the leading geometrical elements (shape) of the particles.

**APPENDIX B: CALCULATION OF SPHERICAL MULTIPOLE COEFFICIENTS**

**1. Numerical approach**

The calculation of spherical multipole coefficients (Eq. 4) is performed with forward spherical harmonic transformation [47,48]. The calculations were performed numerically using numerical library (SHTns), in two consecutive steps. In the first step, the integral over $\varphi$ is performed by calculating the Fourier transform:

$$q_m^\mu(\theta) = \int_0^{2\pi} n_\mu(\theta,\phi) e^{-im\phi} d\phi \qquad (A2)$$

and in the second step we calculate the Legendre transform:

$$q_{lm}^\mu = \frac{r^{l+1}}{R_{\text{eff}}^{l+1}} \int_0^\pi q_m^\mu(\theta) P_l^m(\cos\theta) \sin\theta d\theta. \qquad (A3)$$

The SHTns library uses for the forward spherical harmonic transform data written in spherical coordinates on a sphere, specifically, on nodes of a sphere with discretized latitude $\theta_i$ and longitude $\varphi_i$ $\theta_i$ which are equally spaced along longitudinal coordinate and Gaussian along the latitude. Such distribution of numerical nodes (points) gives more balanced representation of waist region of the sphere in comparison to the poles as if using, e.g. regular grids.

The director filed $n_\mu$ used for determining spherical multipole coefficients is obtained from numerical calculations based on free energy minimization, as explained in Appendix A, and is calculated on discrete points of a cubic mesh, which do not match with points on the sphere. Therefore, we perform trilinear interpolation of the tensor order parameter $Q_{ij}$, to get $Q_{ij}$ at arbitrary point in space:

$$\begin{aligned}Q_{ij}(x,y,z) = &[1-x][1-y][1-z]Q_{ij}(0,0,0) + x[1-y][1-z]Q_{ij}(1,0,0) \\ &+ [1-x]y[1-z]Q_{ij}(0,1,0) + xy[1-z]Q_{ij}(1,1,0) \\ &+ [1-x][1-y]zQ_{ij}(0,0,1) + x[1-y]zQ_{ij}(1,0,1) \\ &+ [1-x]yzQ_{ij}(0,1,1) + xyzQ_{ij}(1,1,1).\end{aligned} \quad (A4)$$

where $(x,y,z)$ with $x, y, z \in [0, 1]$ denotes a location within a selected cube of 8 neighboring points of a square lattice; each point is in the corner of the cube and the corners are labelled with vector of 0 and 1. Performing the calculation of spherical multipole coefficients on the discrete grid, the integral (Eq. A2) reduces to the discrete Fourier transform and the use of the Gauss-Legendre quadrature replaces the integral (Eq. A3) with the sum

$$q_{lm}^{\mu} = \frac{r^{l+1}}{R_{\text{eff}}^{l+1}} \sum_{j=1}^{N_\theta} q_m^{\mu}(\theta_j) P_l^m(\cos\theta_j) w_j, \quad (A5)$$

where $\theta_j$ and $w_j$ are Gauss node angles and Gauss node weights, respectively, and $N_\theta$ is the number of discrete points in latitude.

The radius of the interpolation sphere must be chosen such that the distortions from homogeneous alignment $n_\mu$ are rather small compared to $n_z \approx 1$. We perform the multipole analysis by setting the magnitude of the maximum allowed transversal director field component to be $n_x = 0.1$ on the entire interpolation sphere (see also Fig. 2), where we determine the appropriate interpolation sphere radius $r_i$ with bisection.

For the effective size of the particles (see discussion at the end of Appendix A), we take that the effective radius to be half the length of the particle's dimension in $z$ direction, which can be written (using parameters from Fig. 2) as $R_{\text{eff}} = (2r_a + d_b + r_b)/2$. Note, that we tested various possible selections for the effective radius and for the systems shown, this selection gives most reasonable results; especially, such selection of $R_{\text{eff}}$ accounts rather well for the changes in the geometry of our particles and makes the multipole coefficients comparable in magnitude. Finally, selecting the effective radius, it also defines the center of the composite colloid, which we call the geometrical center and is depicted with red dot in Fig. 2.

## 2. Multipole position

Important parameter in the calculation of the multipolar coefficients is also the position of the interpolation sphere $d_i$, i.e., the actual location of the multipoles. Note that if the distortions of the nematic director are symmetric up-down along the far-field director (in our case along $z$ direction), the location of the multipole is clearly at the mirror plane. Also, if the distortions are rotationally symmetric around $\mathbf{n}_0$, the location of the multipole is along the rotational axis. However, the location of the multipoles and, correspondingly, the choice of the interpolation sphere position become less clear if the distortion - and particle - are asymmetric. In this work we take the interpolation sphere to be centered in the geometric center, which we analyze in more details by varying the center of the interpolation sphere as shown in Figs. 9 and 10.

Figure 9 shows the spherical multipole coefficients $q_{l1}^x$ as a function of the interpolation sphere position $d_i$. As first example, we present a spherical particle with homeotropic anchoring on the surface (Fig. 9a), which has the director structure of an elastic quadrupole, commonly known as the Saturn ring configuration. Figure 9b shows that dipole moment is zero, quadrupole is constant regardless of the position of the interpolation sphere $d_i$, whereas all higher multipole moments are present nonetheless with their magnitudes dependent on $d_i$. Notably, if the center of the interpolation sphere coincides with the geometrical center of the colloidal particle, the multipole coefficients have an extreme or zero, which actually one would expect, and supports the relevance of the geometrical center as the location of the multipoles. Note that the higher multipole moments emerge primarily because the particle is positioned away from the center of the simulation box (in which we calculate the Q tensor profile) and the confinement distorts the exact quadrupolar symmetry of the director field. As second example, a spherical particle with conic degenerate anchoring is presented (Fig. 9c and 9d). The quadrupole coefficient is observed to be constant for all $d_i$; however, other higher multipole coefficients emerge as well and are again dependent on the interpolation sphere position. But, interestingly, again, when the center of the interpolation sphere coincides with the geometrical center of the colloidal sphere at $d_i = -r_a$, all multipoles higher than 16-pole drop to zero. This again supports the relevance of geometric center as a reasonable position for the center of the interpolation sphere in determining the multipole coefficient. Nevertheless, less clearly, Figs. 9e and 9f show the multipole coefficients of a composite colloidal particle with upper sphere radius $r_b = 2/5 r_a$ and at position $d_b = 0$. The quadrupolar moment is constant over all positions of the interpolation sphere, but higher multipoles emerge as well, and notably without clear signature at the geometrical center (such as zero value or maximum/minimum). This result shows that although our use of the geometrical center of particle works well over a range of regimes of particle shapes and anchoring types, as it corresponds exactly to the location of the multipoles, in general, it is only a reasonable approximation. Another example of systems where only geometrical arguments for finding location of multipole centers will fail are (symmetric) colloids (even spheres) with different anchoring strengths (and/or anchoring types) on different parts of the colloidal surface. Overall, this indicates an interesting possible further study, such as if and how, different multipoles of one particle could possibly emerge at different mutually shifted locations.

    Figure 10 shows two chain colloids comprising of two equal spheres joined at the poles. Figure 10a show a pair of colloids with homeotropic anchoring. Two Saturn rings emerge at the waist/equator of each sphere, creating distinct director field deformation of the pattern of the hexadecapole with $n_x > 0$ (see Fig. 1). Nevertheless, as shown by the full analysis (Fig. 10b), the hexadecapolar deformations are also

accompanied by the quadrupolar and weak 64-pole components. Figures 10c and 10d demonstrates another case of a colloidal chain, now with tangential anchoring on the spheres. The nematic profile shows a defect region in the neck of the particle and boojums at each free pole. The corresponding director field resembles the hexadecapole $n_x$<0 (as shown in Fig. 1), effectively, somewhat stretched along $z$ axis, which turns out excites other (symmetrical along $z$-axis) multipoles. For this case, the strongest multipole is the hexadecapole, followed by the quadrupole and 64-pole, whereas other multipoles are zero in the geometrical center of the composite chain particle, as conditioned by the symmetry of the particle distortion.

**APPENDIX C: POLARIMETRIC IMAGING OF COLLOIDS IN LIQUID CRYSTALS**

In addition to the standard technique of polarizing optical microscopy, we used polarimetric imaging of structures around colloidal particles with measurements of the parameters of polarized light emerging from the sample on a pixel-by-pixel basis. To determine the orientation $\chi$ and ellipticity $e$ of the light's polarization ellipse after traversing the nematic sample with a colloidal inclusion (Fig. 11a,b), we used the rotating quarter-wave-plate (QWP) measurements [64]. The measurement setup is shown in Fig. 11c, where we used a narrow band filter with central wavelength at 546 nm after a halogen lamp as a light source. The light incident on the sample was polarized with a linear polarizer. In the optical path, the QWP is inserted after the sample and is followed by an analyzer fixed along $x$-axis. This setup allows for the measurements of polarization ellipse parameters of light passing the sample by using intensities of light transmitted through the system polarizer-sample-QWP-analyzer at different QWP orientations. The QWP can be rotated by an angle $\theta$ with respect to analyzer direction and Stokes parameters (Fig. 11b) can be found as follows [64]:

$$S_0 = Z_1 - Z_3, \quad S_1 = 2Z_3, \quad S_2 = 2Z_4, \quad S_3 = Z_2 \tag{C1}$$

where coefficients *A*, *B*, *C*, *D* are given by

$$Z_1 = \frac{2}{N_p}\sum_{i=1}^{N_p} I_i, \quad Z_2 = \frac{4}{N_p}\sum_{i=1}^{N_p} I_i \sin 2\theta_{pi}, \quad Z_3 = \frac{4}{N_p}\sum_{i=1}^{N_p} I_i \cos 4\theta_{pi}, \quad Z_4 = \frac{4}{N_p}\sum_{i=1}^{N_p} I_i \sin 4\theta_{pi}, \quad (C2)$$

where $N_p$ is a number of angels $\theta_{pi}$ at which the intensity $I_i$ of transmitted light was measured. We measured the intensity of transmitted light at the orientation of the fast axis of QWP with respect to an analyzer from $\theta_p=0°$ to $\theta_p=180°$ with a step of 22.5°. Following this procedure, polarization ellipse parameters $\chi$ and $e$ can be determined from expressions $\tan 2\chi = S_2/S_1$ and $\sin 2e = S_3/S_0$. The intensity of transmitted light after an analyzer corresponding to each pixel was recorded with a CCD camera. A large matrix of intensities corresponding to pixels of camera was recorded for each $\theta_p$ and polarization parameters were calculated for each pixel, yielding polarimetric images of colloidal particles and distortions around them (Fig. 7o). As shown using the examples of dimer composite colloidal particles, the polarimetric imaging results are consistent with polarizing micrographs and numerically calculated director structures, revealing how different multipoles can be induced by composite colloidal objects that we study.

**References**

[1] W. Poon, "Colloids as Big Atoms," Science **304**, 830 (2004).
[2] G.-R. Yi, D. J. Pine, and S. Sacanna, "Recent Progress on Patchy Colloids and Their Self-Assembly," J. Phys.: Condens. Matter **25**, 193101 (2013).
[3] D. Frenkel, "Playing Tricks with Designer Atoms," Science **296**, 65 (2002).
[4] W. C. K. Poon, "Colloids as Big Atoms: the Genesis of a Paradigm," J. Phys. A: Math. Theor. **49**, 401001 (2016).
[5] W. B. Rogers, W. M. Shih, and V. N. Manoharan, "Using DNA to program the selfassembly of colloidal nanoparticles and microparticles," Nat. Rev. Mater. **1**, 1 (2016).

[6] S. Sacanna, M. Korpics, K. Rodriguez, L. Colón-Meléndez, S.-H. Kim, D. J. Pine, and G.-R. Yi, "Shaping Colloids for Self-Assembly," Nat. Commun. **4**, 1688 (2013).

[7] Y. Wang, Y. Wang, D. R. Breed, V. N. Manoharan, L. Feng, A. D. Hollingsworth, M. Weck, and D. J. Pine, "Colloids with Valence and Specific Directional Bonding," Nature **491**, 51 (2012).

[8] V. N. Manoharan, M. T. Elsesser, and D. J. Pine, "Dense Packing and Symmetry in Small Clusters of Microspheres," Science **301**, 483 (2003).

[9] É. Ducrot, M. He, G.-R. Yi, and D. J. Pine, "Colloidal Alloys with Preassembled Clusters and Spheres," Nat. Mater. **16**, 652 (2017).

[10] D. R. Nelson, "Toward a Tetravalent Chemistry of Colloids," Nano Lett. **2**, 1125 (2002).

[11] K.-T. Wua, L. Feng, R. Sha, R. Dreyfus, A. Y. Grosberg, N. C. Seeman, and P. M. Chaikin, "Polygamous Particles," Proc. Natl. Acad. Sci. U.S.A. **109**, 18731 (2012).

[12] D. Nykypanchuk, M. M. Maye, D. van der Lelie, and O. Gang, "DNA-Guided Crystallization of Colloidal Nanoparticles," Nature **451**, 549 (2008).

[13] P. L. Biancaniello, A. J. Kim, and J. C. Crocker, "Colloidal Interactions and Self-Assembly Using DNA Hybridization," Phys. Rev. Lett. **94**, 058302 (2005).

[14] Y. Wang, Y. Wang, X. Zheng, É. Ducrot, J. S. Yodh, M. Weck, and D. J. Pine, "Crystallization of DNA-Coated Colloids," Nat. Commun. **6**, 7253 (2015).

[15] I. I. Smalyukh, "Liquid Crystal Colloids," Annu. Rev. Condens. Matter Phys. **9**, 207 (2018).

[16] P. Poulin, H. Stark, T. C. Lubensky, and D. A. Weitz, "Novel Colloidal Interactions in Anisotropic Fluids," Science **275**, 1770 (1997).

[17] M. Škarabot, M. Ravnik, S. Žumer, U. Tkalec, I. Poberaj, D. Babič, N. Osterman, and I. Muševič, "Interactions of Quadrupolar Nematic Colloids," Phys. Rev. E **77**, 031705 (2008).

[18] C. P. Lapointe, T. G. Mason, and I. I. Smalyukh, "Shape-Controlled Colloidal Interactions in Nematic Liquid Crystals," Science **326**, 1083 (2009).

[19] R. W. Ruhwandl and E. M. Terentjev, "Long-Range Forces and Aggregation of Colloid Particles in a Nematic Liquid Crystal," Phys. Rev. E **55**, 2958 (1997).

[20] U. Ognysta, A. Nych, V. Nazarenko, I. Muševič, M. Škarabot, M. Ravnik, S. Žumer, I. Poberaj, and D. Babič, "2D Interactions and Binary Crystals of Dipolar and Quadrupolar Nematic Colloids," Phys. Rev. Lett. **100**, 217803 (2008).

[21] B. Senyuk, Q. Liu, S. He, R. D. Kamien, R. B. Kusner, T. C. Lubensky, and I. I. Smalyukh, "Topological Colloids," Nature **493**, 200 (2013).

[22] T. C. Lubensky, D. Pettey, N. Currier, and H. Stark, "Topological Defects and Interactions in

Nematic Emulsions," Phys. Rev. E **57**, 610 (1998).

[23] P. Poulin and D. A. Weitz, "Inverted and Multiple Nematic Emulsions," Phys. Rev. E **57**, 626 (1998).

[24] U. M. Ognysta, A. B. Nych, V. A. Uzunova, V. M. Pergamenschik, V. G. Nazarenko, M. Škarabot, and I. Muševič, "Square Colloidal Lattices and Pair Interaction in a Binary System of Quadrupolar Nematic Colloids," Phys. Rev. E **83**, 041709 (2011).

[25] I. Muševič, "Optical Manipulation and Self-assembly Of Nematic Colloids: Colloidal Crystals and Superstructures," Liq. Cryst. Today **19**, 2 (2010).

[26] V. M. Pergamenshchik and V. A. Uzunova, "Dipolar Colloids In Nematostatics: Tensorial Structure, Symmetry, Different Types, and Their Interaction," Phys. Rev. E **83**, 021701 (2011).

[27] O. M. Tovkach, S. B. Chernyshuk, and B. I. Lev, "Theory of Elastic Interaction Between Arbitrary Colloidal Particles in Confined Nematic Liquid Crystals," Phys. Rev. E **86**, 061703 (2012).

[28] V. A. Uzunova and V. M. Pergamenshchik, "Chiral Dipole Induced by Azimuthal Anchoring on The Surface of a Planar Elastic Quadrupole," Phys. Rev. E **84**, 031702 (2011).

[29] V. M. Pergamenshchik, "Elastic Multipoles in the Field of the Nematic Director Distortions," Eur. Phys. J. E **37**, 387 (2014).

[30] O. O. Ramdane, Ph. Auroy, S. Forget, E. Raspaud, Ph. Martinot-Lagarde, and I. Dozov, "Memory-free Conic Anchoring of Liquid Crystals on a Solid Substrate," Phys. Rev. Lett. **84**, 3871 (2000).

[31] B. Senyuk, O. Puls, O. M. Tovkach, S. B. Chernyshuk, and I. I. Smalyukh, "Hexadecapolar colloids," Nat. Commun. **7**, 10659 (2016).

[32] R. E. Raab and O. L. De Lange, *Multipole Theory in Electromagnetism: Classical, Quantum, and Symmetry Aspects, with Applications* (Oxford University Press, Oxford, U.K., 2004).

[33] J. Petschulat, C. Menzel, A. Chipouline, C. Rockstuhl, A. Tünnermann, F. Lederer, and T. Pertsch, "Multipole Approach to Metamaterials," Phys. Rev. A **78**, 32 (2008).

[34] N. Papasimakis, V. A. Fedotov, V. Savinov, T. A. Raybould, and N. I. Zheludev, "Electromagnetic Toroidal Excitations in Matter and Free Space," Nat. Mater. **15**, 263 (2016).

[35] J. D. Jackson, *Classical Electrodynamics* (Wiley, New York, 1998) 3rd ed.

[36] E. M. Purcell, *Electricity and Magnetism* (Cambridge University Press, 2011) 2nd ed.

[37] M. Born, *Atomic Physics* (Dover Publications, 1989) 8th ed.

[38] P. G. de Gennes and J. Prost, *The Physics of Liquid Crystals* (Oxford University Press, Oxford, 1993).

[39] J. W. Kim, R. J. Larsen, and D. A. Weitz, "Synthesis of Nonspherical Colloidal ParticleswWith

Anisotropic Properties," J. Am. Chem. Soc. **128**, 14374-14377 (2006).

[40] F. Ma, S. Wang, L. Smith, and N. Wu, "Two-Dimensional Assembly of Symmetric Colloidal Dimers under Electric Fields," Adv. Funct. Mater. **22**, 4334-4343 (2012).

[41] S. Wang, F. Ma, H. Zhao, and N. Wu, "Bulk Synthesis of Metal-Organic Hybrid Dimers and Their Propulsion under Electric Fields," ACS Appl. Mater. Interfaces **6**, 4560-4569 (2014).

[42] M. C. M. Varney, Q. Zhang, M. Tasinkevych, N. M. Silvestre, K. A. Bertness, and I. I. Smalyukh, "Periodic Dynamics, Localization Metastability, and Elastic Interaction of Colloidal Particles with Confining Surfaces and Helicoidal Structure of Cholesteric Liquid Crystals," Phys. Rev. E **90**, 062502 (2014).

[43] M. C. M. Varney, N. J. Jenness, and I. I. Smalyukh, "Geometrically unrestricted, topologically constrained control of liquid crystal defects using simultaneous holonomic magnetic and holographic optical manipulation," Phys. Rev. E **89**, 022505 (2014).

[44] R. P. Trivedi, T. Lee, K. Bertness and I. I. Smalyukh, "Three Dimensional Optical Manipulation and Structural Imaging of Soft Materials by Use of Laser Tweezers and Multimodal Nonlinear Microscopy," Opt. Express **18**, 27658-27669 (2010).

[45] R. P. Trivedi, D. Engström, and I. I. Smalyukh, "Optical Manipulation of Colloids and Defect Structures in Anisotropic Liquid Crystal Fluids," J. Opt. **13**, 044001 (2011).

[46] M. Ravnik and S. Žumer, "Landau-de Gennes Modelling of Nematic Liquid Crystal Colloids," Liq. Cryst. **36**, 1201 (2009).

[47] N. Schaeffer, "Efficient Spherical Harmonic Transforms Aimed at Pseudospectral Numerical Simulations," Geochem. Geophys. Geosyst. 14, 751 (2013).

[48] N. Schaeffer, SHTns. https://users.isterre.fr/nschaeff/SHTns/, Accessed: 10. 5. 2018.

[49] M. Frigo and S. G. Johnson, "The Design and Implementation of FFTW3," Proc. IEEE **93**, 216 (2005).

[50] M. Frigo and S. G. Johnson, FFTW (2018). http://www.fftw.org/, Acessed: 10. 5. 2018.

[51] B. Senyuk, Q. Liu, P. D. Nystrom and I. I. Smalyukh, "Repulsion-Attraction Switching of Nematic Colloids Formed by Liquid Crystal Dispersions of Polygonal Prisms," Soft Matter **13**, 7398-7405 (2017).

[52] A. Martinez, M. Ravnik, B. Lucero, R. Visvanathan, S. Žumer, and I. I. Smalyukh, "Mutually Tangled Colloidal Knots and Induced Defect Loops in Nematic Fields," Nat. Mater. **13**, 258-263 (2014).

[53] A Martinez, L. Hermosillo, M. Tasinkevych, and I. I. Smalyukh, "Linked Topological Colloids in a

Nematic Host," Proc. Natl. Acad. Sci. U.S.A. **112**, 4546-4551 (2015).

[54] M. Conradi, M. Ravnik, M. Bele, M. Zorko, S. Žumer, I. Muševič, "Janus Nematic Colloids," Soft Matter **5**, 3905-3912 (2009).

[55] H. Mundoor, B. Senyuk, and I. I. Smalyukh, "Triclinic Nematic Colloidal Crystals from Competing Elastic and Electrostatic Interactions," Science **352**, 69-73 (2016).

[56] H. Mundoor, S. Park, B. Senyuk, H. H. Wensink, and I. I. Smalyukh, "Hybrid Molecular-Colloidal Liquid Crystals," Science **360**, 768-771 (2018).

[57] A. Mertelj, D. Lisjak, M. Drofenik, and M. Čopič, "Ferromagnetism in Suspensions of Magnetic Platelets in Liquid Crystal," Nature **504**, 237-241 (2013).

[58] Q. Liu, P. J. Ackerman, T. C. Lubensky, and I. I. Smalyukh, "Biaxial Ferromagnetic Liquid Crystal Colloids," Proc. Natl. Acad. Sci. U.S.A. **113**, 10479-10484 (2016).

[59] Q. Liu, Y. Yuan and I. I. Smalyukh, "Electrically and Optically Tunable Plasmonic Guest–Host Liquid Crystals with Long-Range Ordered Nanoparticles," Nano Lett. **14**, 4071-4077 (2014).

[60] P. J. Ackerman, H. Mundoor, I. I. Smalyukh, and J. van de Lagemaat, "Plasmon–Exciton Interactions Probed using Spatial Coentrapment of Nanoparticles by Topological Singularities," ACS Nano **9**, 12392-12400 (2015).

[61] A. Bregar, T. J. White, and M. Ravnik, "Refraction of Light on Flat Boundary of Liquid Crystals or Anisotropic Metamaterials," Liq. Cryst. Rev. **5**, 53-68 (2017).

[62] M. Nobili and G. Durand, "Disorientation-Induced Disordering at a Nematic-Liquid-Crystal-Solid Interface," Phys. Rev. A **46**, R6174 (1992).

[63] H. Stark, "Physics of Colloidal Dispersions in Nematic Liquid Crystals," Phys. Rep. **351**, 387 (2001).

[64] B. Schaefer, E. Collett, R. Smyth, D. Barrett, and B. Fraher, "Measuring the Stokes Polarization Parameters," Am. J. Phys. **75**, 163 (2007).

**Figures**

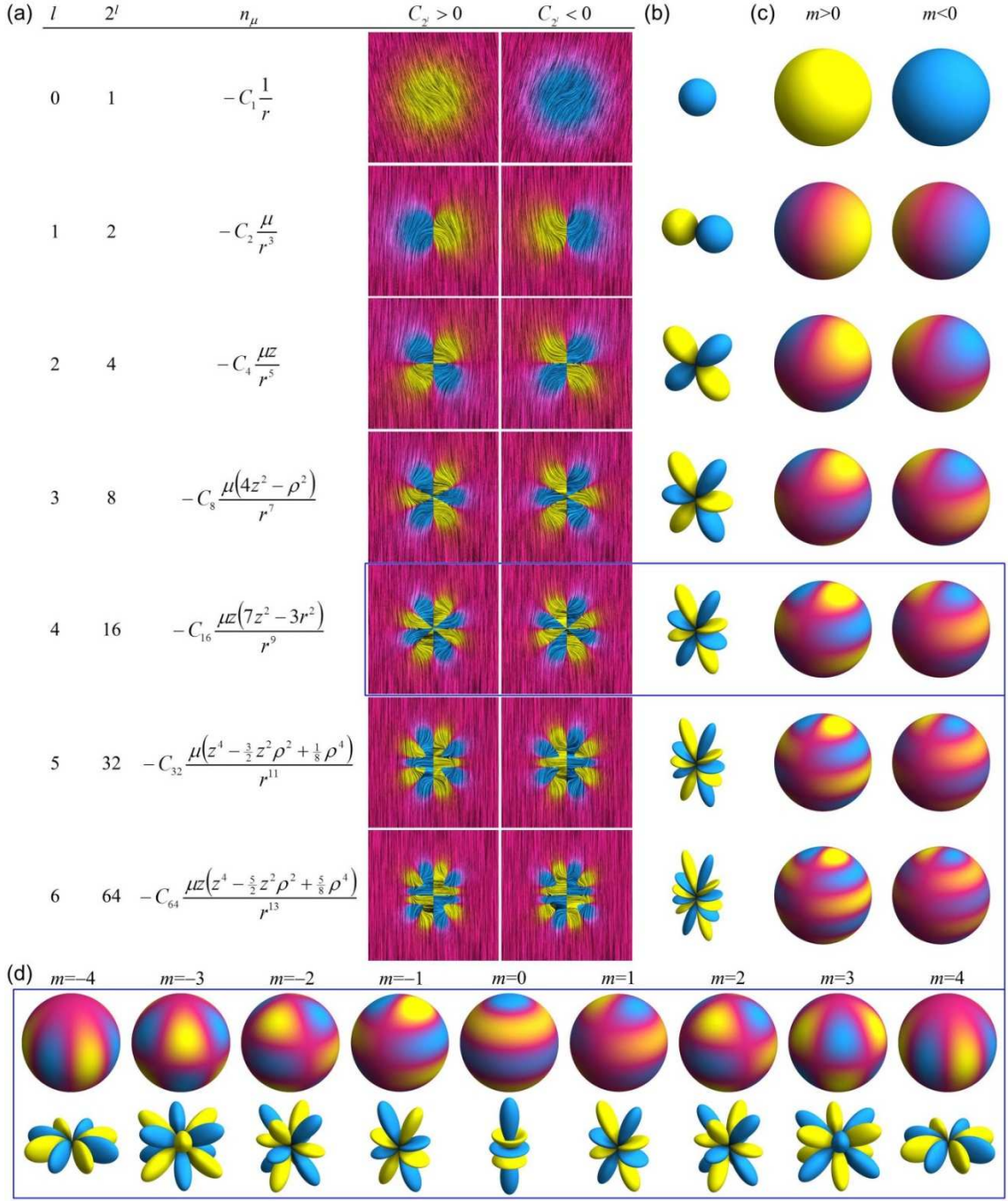

FIG. 1. Correspondence between elastic multipoles, electrostatic charge distributions on spheres and atomic orbitals of elements. (a) Analytic elastic multipoles. The *xz* cross section of the normalized director field ($n_x$, $n_y$, 1) of individual multipoles in Cartesian coordinates. Note that $r^2 = x^2 + y^2 + z^2$ and $\rho^2 = x^2 + y^2$. (b) Diagrams for *s*-, *p*-, *d*-, *f*-, *g*-, *i*- and *k*- atomic orbitals calculated using the angular wavefunction. (c) Elastic multipoles around spherical particles with a tilt of director at their surface. (d) Elastic multipoles (top row) and atomic orbitals (bottom row) for hexadecapoles with different *m* values.

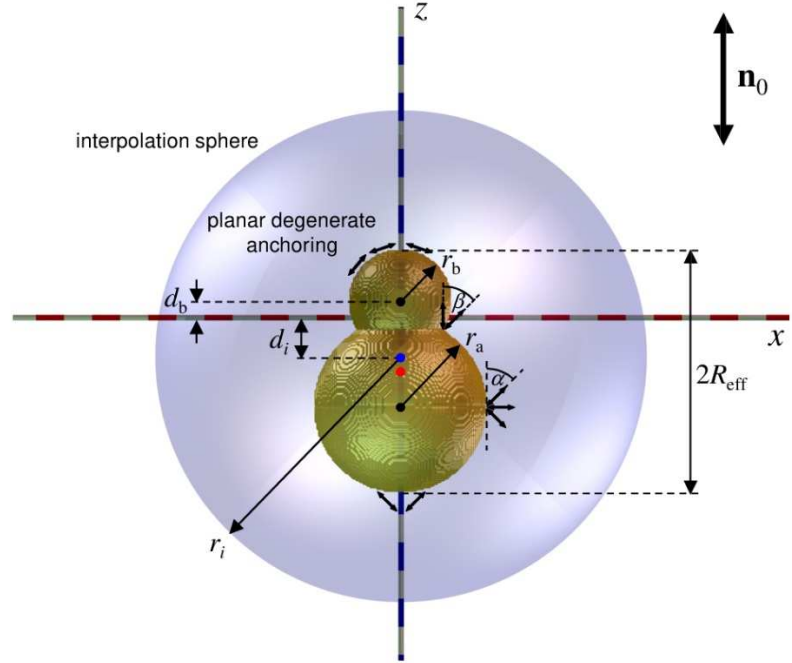

FIG. 2. Model of a colloidal particle consisting of two spheres used in calculations, where the lower sphere has a constant radius $r_a$ and is positioned at a distance - $r_a$ from the origin. The upper sphere's radius $r_b$ and position $d_b$ vary in the process of analysis. The anchoring on the upper sphere is planar degenerate, while that on the lower sphere it is conic degenerate with a tilt angle $\alpha$. At the neck, where the two spheres with distinct anchoring meet, the angle between the two anchoring directions is represented with angle $\beta$. Blue sphere shows the interpolation sphere of radius $r_i$ and blue filled circle depicts its center displaced by $d_i$ from the origin. Red filled circle represents geometrical center of the composite colloidal particle whereas half of the composite particle length along $z$-axis represents the effective particle radius $R_{eff}$.

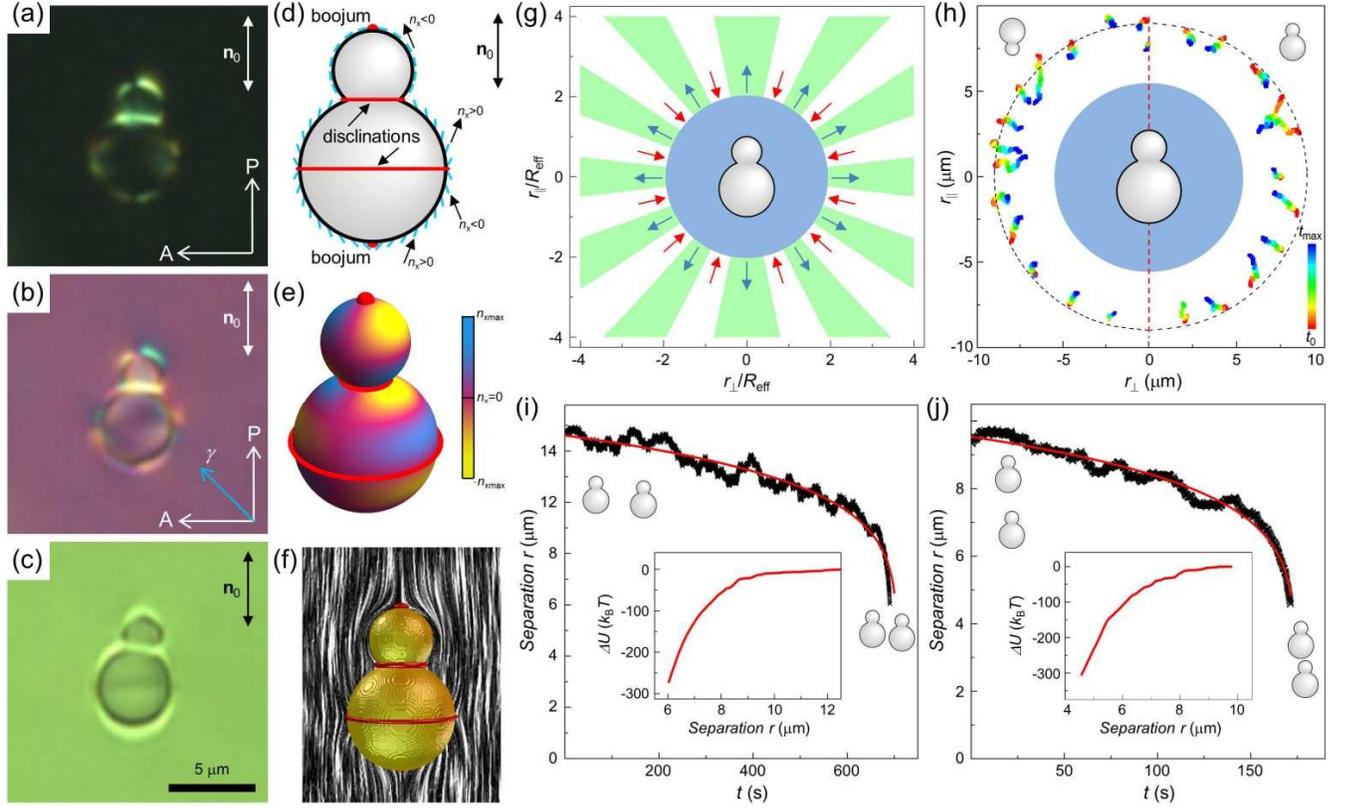

FIG. 3. Elastic multipole induced by a gourd-shaped particle. (a-c) Textures of a sample with a gourd-like particle in a nematic cell from optical polarizing, without (a) and with (b) a retardation plate, and bright field (c) microscopy. (d,e) Schematic diagram (blue lines) of **n(r)** at the surface and corresponding color-coded diagram of the $n_x$. (f) Calculated **n(r)**. (g, h) Calculated for elastic 64-poles with only 64-pole non-zero coefficient (g) and experimentally measured (h) map of elastic interactions between gourd particles with respect to the director. Dashed red line in (h) separates maps of interactions for parallel (on the right) and antiparallel particles (on the left). (i,j) Separation vs. time dependence for gourd particles elastically interacting along the direction approximately perpendicular (i) and parallel (j) to $\mathbf{n}_0$. Insets show the corresponding dependence of the interaction potential from the distance between particles with a 64-pole coefficient $b_6 \approx -3 \cdot 10^{-5}$ extracted from fitting.

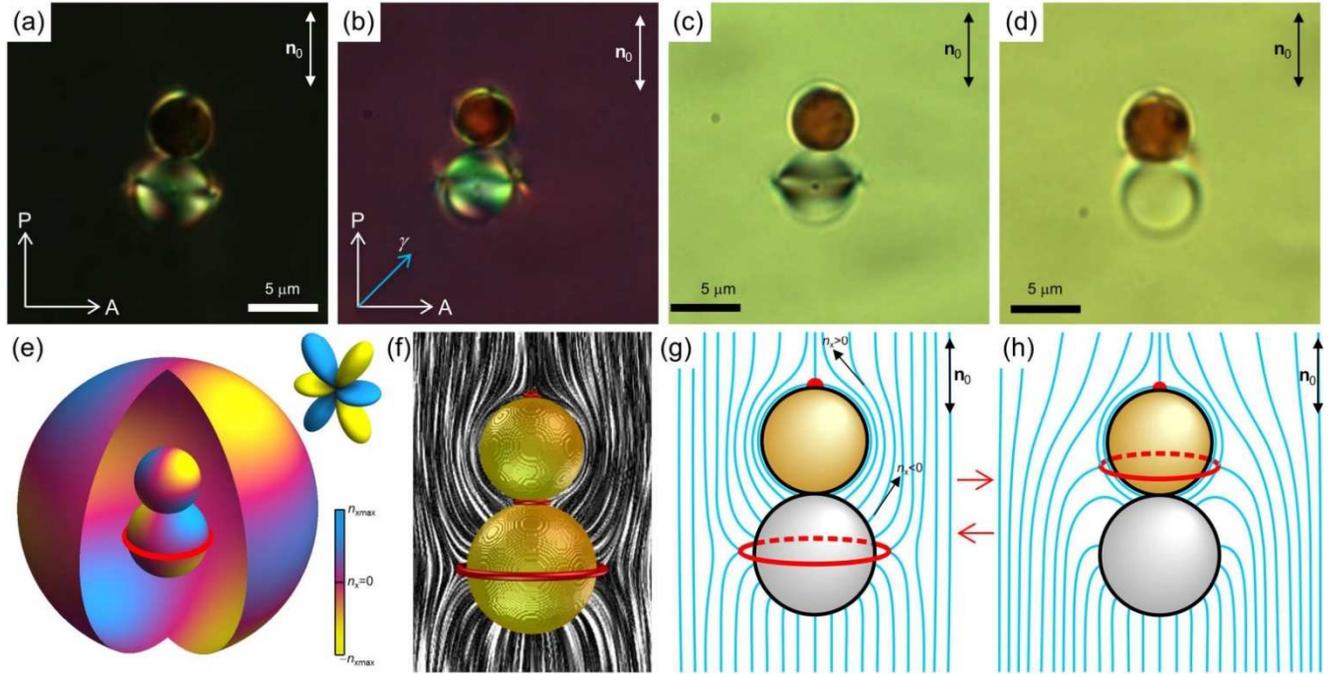

FIG. 4. Elastic multipoles induced by a pair of dissimilar particles, one (top) with planar and another (bottom) with homeotropic anchoring. (a-d) Textures of a sample with particles in a nematic cell obtained by optical polarizing, with (b) and without (a) a retardation plate, and bright field (c,d) microscopy. (e-g) Schematic diagram (blue lines) of **n**(**r**) (g) also calculated in f and (e) corresponding colour-coded diagram of the $n_x$ directly at the surface of the particle shown in a-c and at the surface of the interpolation sphere. Inset shows a corresponding atomic orbital. (h) Schematic diagram of **n**(**r**) around a pair of particles shown in d.

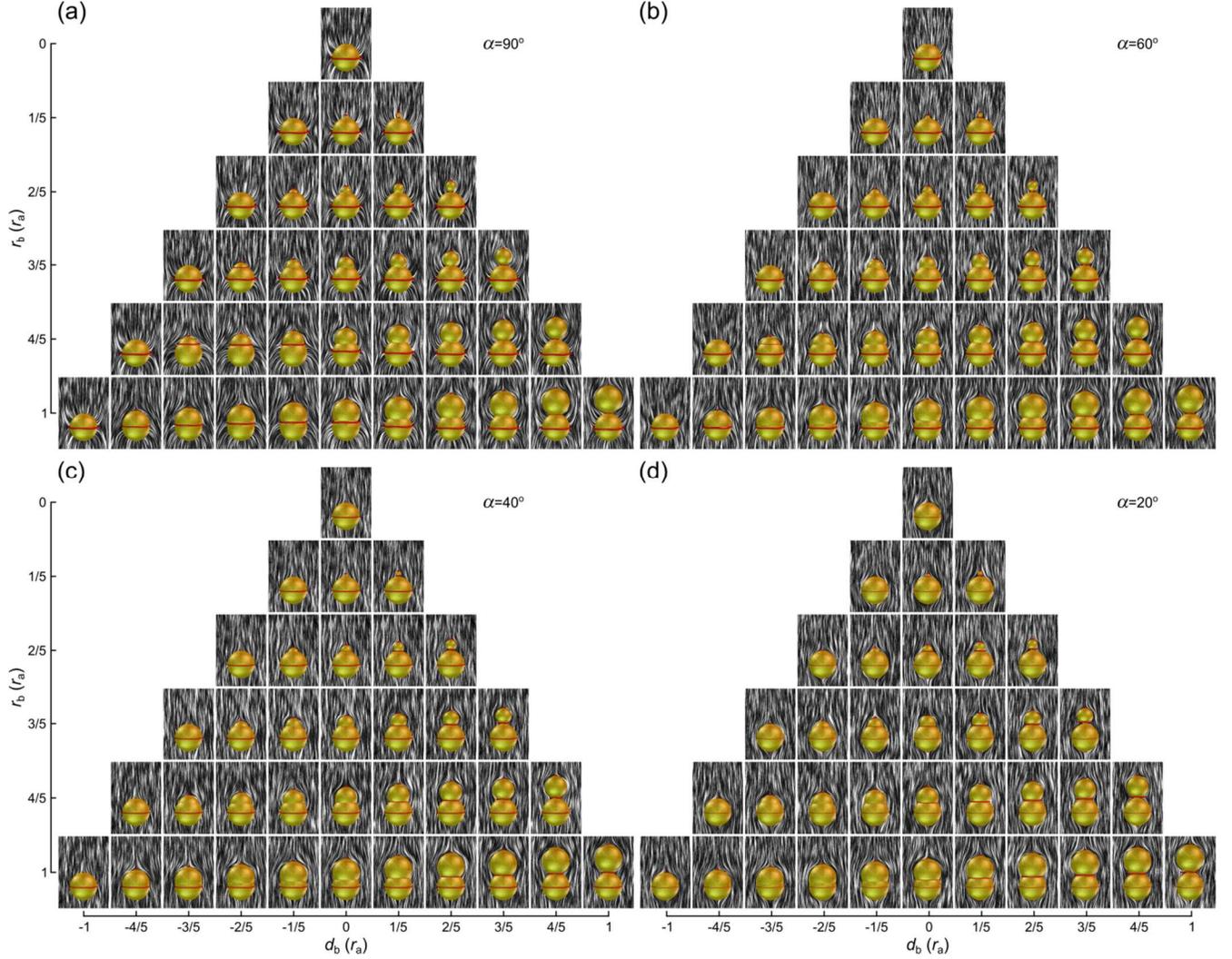

FIG. 5. Composite colloids with planar degenerate anchoring (upper sphere) and tilted anchoring with various angles $\alpha$ (lower sphere) and their corresponding director field structures. Pyramid-like diagram shows the composite colloidal particles (yellow isosurfaces) with variable radius $r_b$ and position $d_b$ of the upper sphere and conic anchoring at angle $\alpha = 90°$ (a), $\alpha = 60°$ (b), $\alpha = 40°$ (c) and $\alpha = 20°$ (d) on the lower sphere. Defect regions are indicated with red isosurfaces ($S = 0.44$), whereas the director field is illustrated with black and white streaks. (a) Defects generally emerge at the lower sphere equator, upper sphere north pole and possibly at the neck where both spheres intersect. The Saturn ring surrounding the lower sphere can move upwards or even to the neck, joining with the neck defect, (lower left side of the pyramid). (b, c) Defects generated by conic anchoring emerge precisely at the equator of the lower sphere, whereas planar degenerate anchoring induces boojum defect at the north pole of the upper sphere. The mismatch between the two anchoring directions forms the surface neck defect. The profile of the neck defect is varied by changing the upper sphere geometry and position; it completely vanishes for composite colloids on the third diagonal from the left side of the pyramid. (d) The defects generated by conic

anchoring emerge precisely at the equator of the lower sphere, whereas planar degenerate anchoring induces boojum defect at the north pole of the upper sphere. The mismatch between the two anchoring directions forms the surface neck defect. The strength of the neck defect is varied by changing the upper sphere geometry and position, it completely vanishes for composite colloids on the first diagonal from the left side of the pyramid.

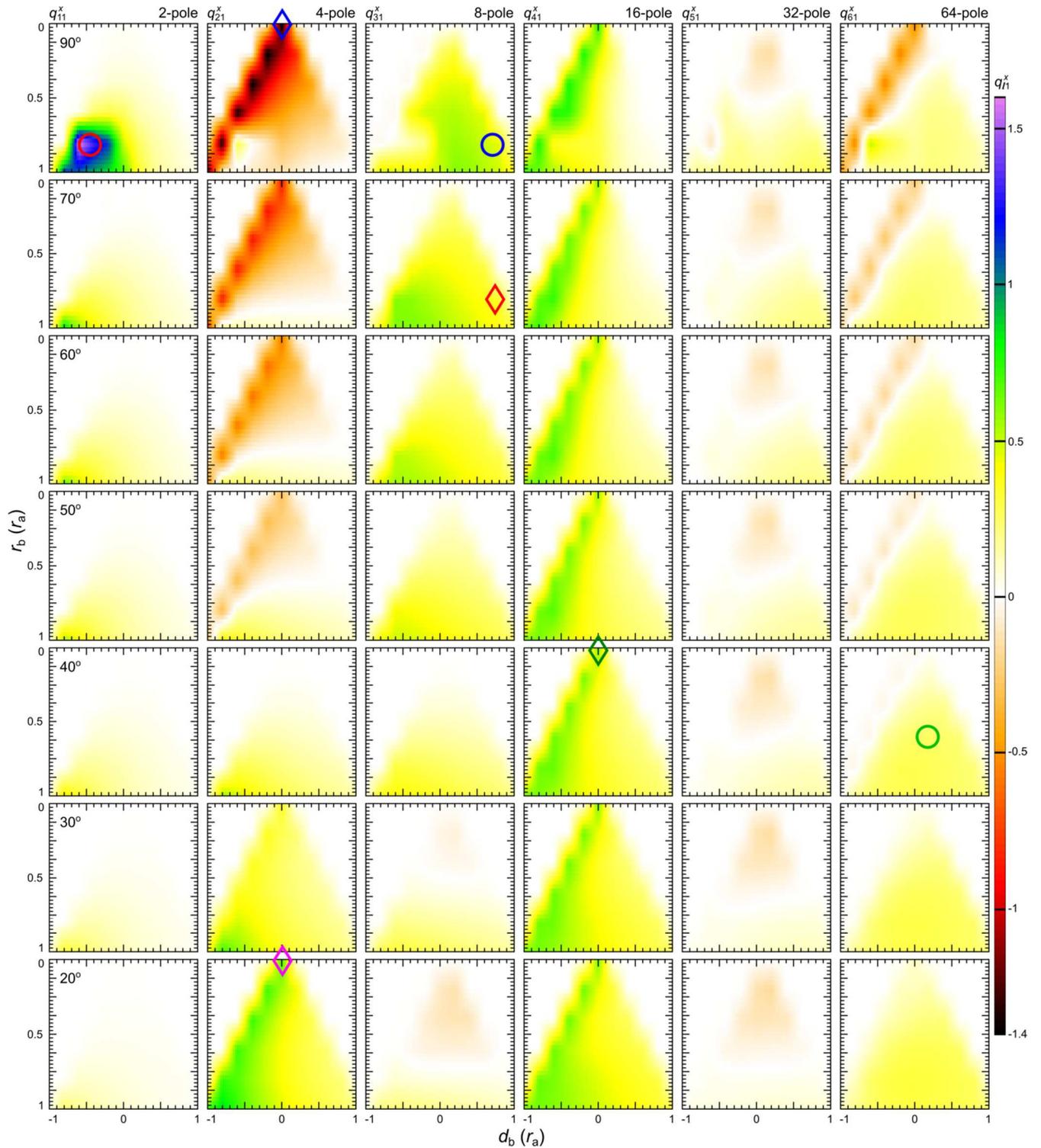

FIG. 6. Spherical multipole moments $q_{l1}^{x}$ ($l = 1$-$6$) calculated at the particles geometrical center for all simulated composite gourd-like colloids with various conic anchoring angles at the lower lobe. Strong dipole emerges for particles near $r_b = 4/5$, $d_b = -2/5$, quadrupole is most pronounced for spherical particle with homeotropic or nearly tangential boundary condition. Octupole is larger for composites with large

upper spheres and conic anchoring, whereas the strong hexadecapole emerges for spherical particles with conic anchoring on the surfaces. 32- and 64-pole generally weak compared to the others because the neck defect region only slightly perturb the director field. Note that the multipole moments are calculated only in discrete points and interpolated for better visibility. Circles point to multipole moments of configurations with parameters close to experimentally realized: red circle corresponds to a dipole in Fig. 4d,h, blue circle corresponds to the colloid with an octupolar configuration in Fig. 4a and green circle corresponds to a 64-pole in Fig. 3. Blue rhombs point to dominant or "pure" multipole moments: blue and magenta rhombs show respectively quadrupoles with a Saturn ring and boojums; red and green rhombs point, respectively, to dominant octupole and hexadecapole.

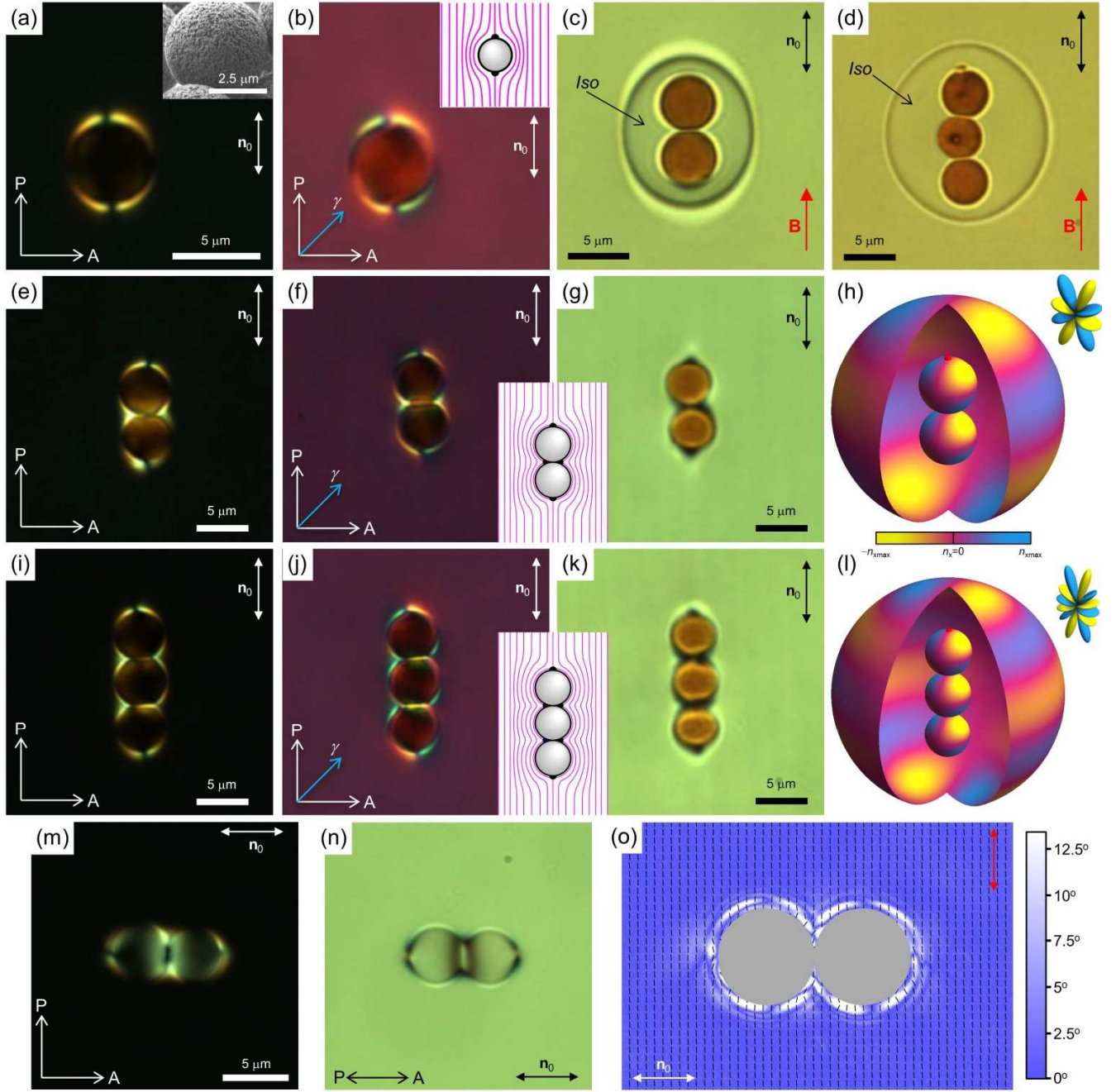

FIG. 7. Elastic multipoles around a chain of similar particles with planar anchoring. (a-g, i-k) Textures of particles in a nematic cell from optical polarizing, with (b,f,j) and without (a,e,i) a retardation plate, and bright field (c,d,g,k) microscopy. P and A are crossed polarizer and analyzer and $\gamma$ is a slow axis of a full wave (530 nm) retardation plate. **B** is magnetic field. Inset in (a) shows the SEM image of the epoxy particles. Insets show a schematic diagram of the director field **n(r)** (magenta lines) around epoxy particles. (h, l) Colour-coded diagrams of the $n_x$ component of **n(r)**, which is caused by its tilt away from **n**$_0$, directly at the surface of the particles and at the surface of the interpolation sphere. (o) Polarimetric micrograph of spatial distribution of polarization state of the green imaging light passing through the sample shown in m and n and polarized in the direction shown by a red arrow. Thin black bars indicate the orientation of the polarization ellipses and background color shows the distribution of their ellipticity (Appendix C) according to the color-coded bar on the right.

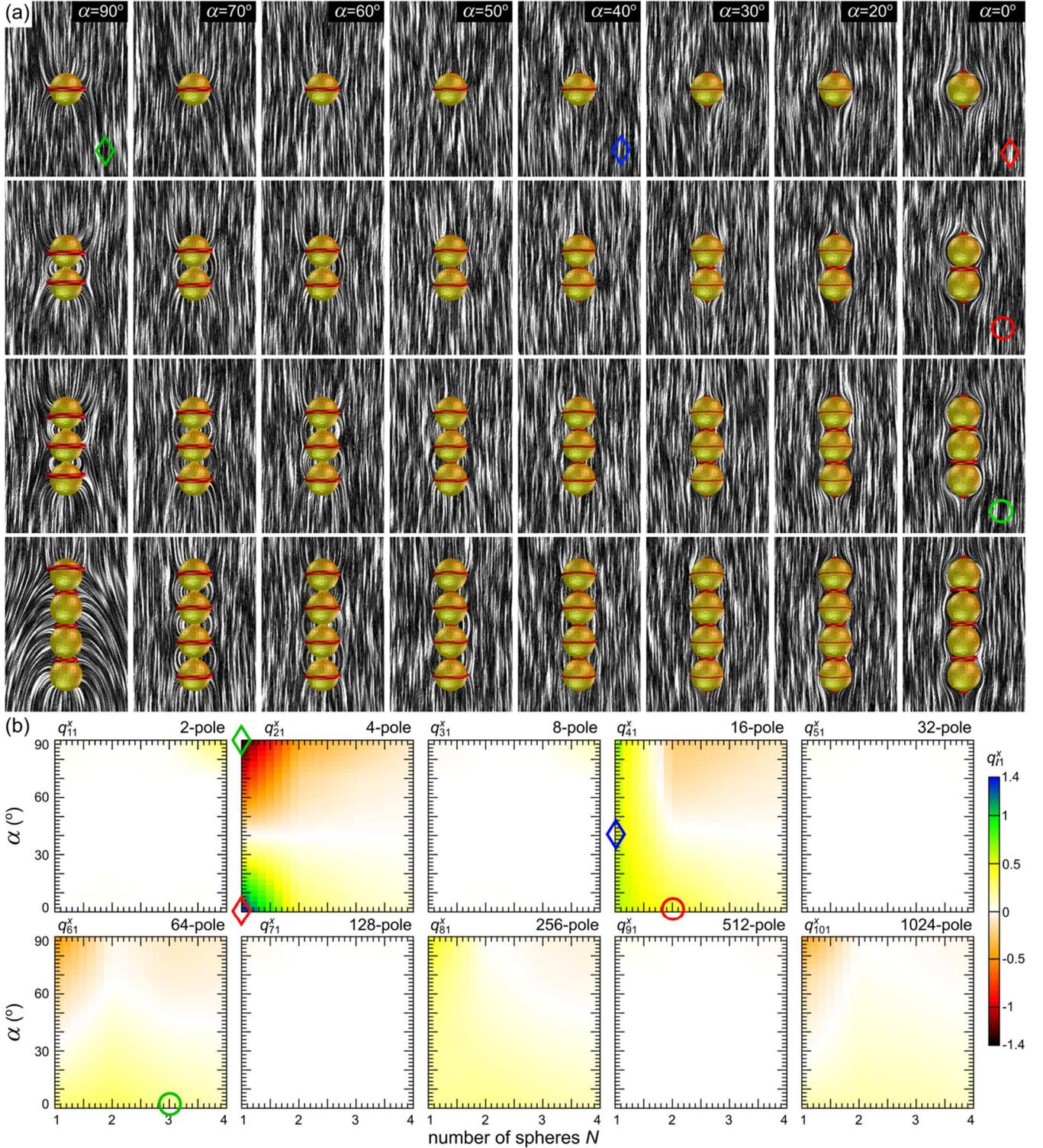

FIG. 8. Chain particles with conic anchoring and their elastic multipole moments. (a) The defects transitions from Saturn ring on the equator (homeotropic anchoring) to combination of both Saturn rings, neck defects and boojums (moderate angle $\alpha$) and ultimately to combination of neck defects and boojums. The surface anchoring tilt angle notably affects the director field distortion around the particle and consequently its multipole moment. (b) Chain composites have mirror symmetry along $z$-axis, therefore multipole moments with odd $l$ are not present. Circles show configurations (a) and parameter space (b) also realized in the experiments (see Fig. 7). Rhombs show configurations (a) and parameter space (b) with pure multipoles. Strong quadrupole emerges for spheres with tangential or homeotropic anchoring on the surface (marked respectively by green and red rhombs). Sphere with conic anchoring angle $\alpha = 40°$ express dominant hexadecapole (marked by a blue rhomb) since the quadrupole is absent. The hexadecapolar moment is also dominant in the particles comprising of two spheres with tangential anchoring (marked by a red circle). 64-pole is weak, but for the particle comprising of three spheres it is the most prominent multipole (marked by a green circle).

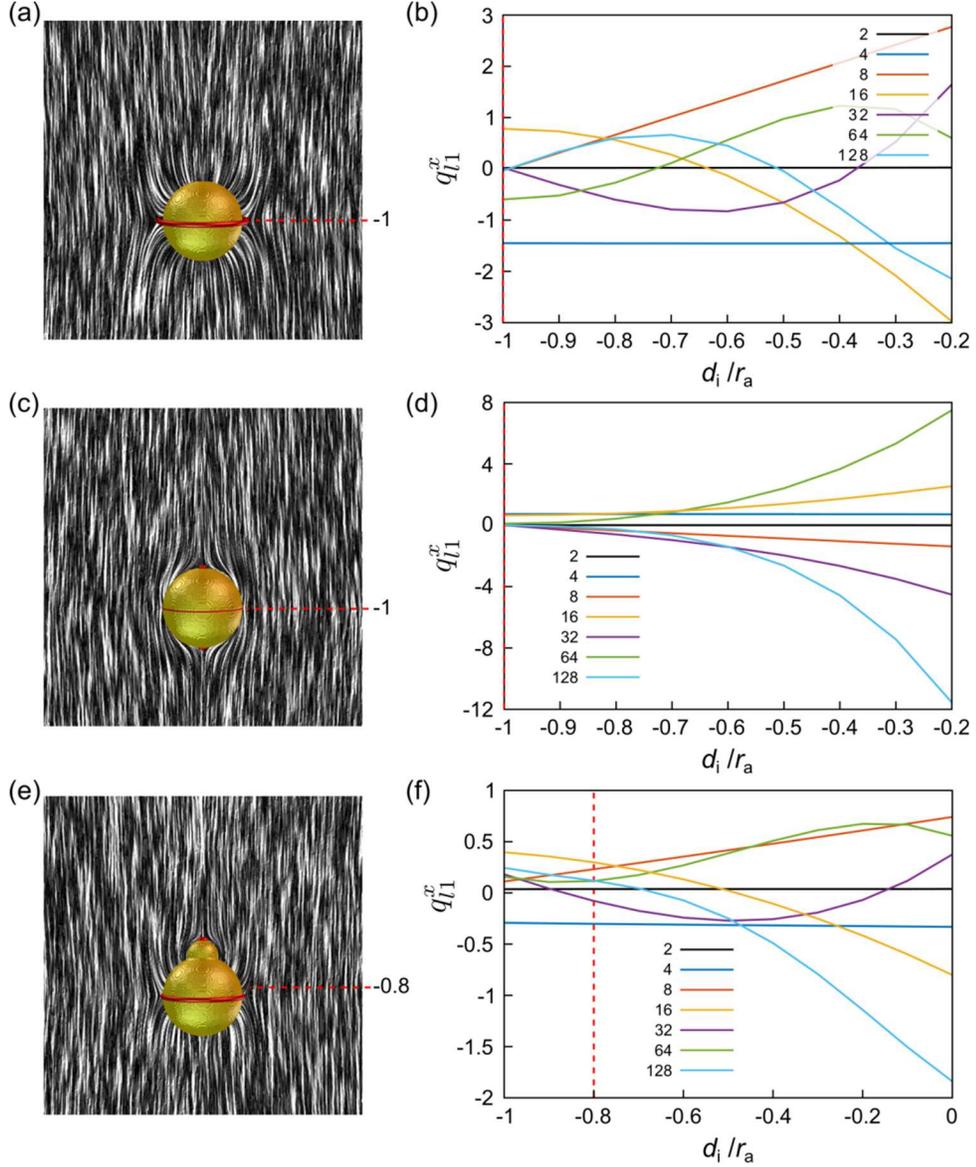

FIG. 9. Spherical multipole coefficient $q_{l1}^x$ of selected simulated composite nematic colloids: (a) Colloidal particle with homeotropic boundary conditions centered at $r_a$ below the origin induces a quadrupolar distortion of the director field, commonly known in the literature as standard Saturn ring configuration. (b) Corresponding graph shows constant quadrupole coefficient at all displacements $d_i$, other multipoles have extreme or zero at $d_i = -r_a$, corresponding to the geometrical center of the spherical colloid. (c) Spherical colloidal particle with conic anchoring at angle $\alpha = 20°$ centered at $r_a$ below the origin. (d) Plot shows that quadrupolar coefficient is constant at all $d_i$, whereas other coefficients higher then 16-pole are zero when center of the interpolation sphere coincides with geometrical center $d_i = d_a$. (e) Composite colloidal particle at $r_a$ below the origin with $d_b = 0$, $r_b = 2r_a/5$ and conic anchoring at angle $\alpha = 60°$. (f) Quadrupolar coefficient is constant regardless the position of the interaction sphere, whereas higher

multipole moments show complex variations even at the geometrical center. Geometrical centers of the composite colloids are depicted with red dashed line.

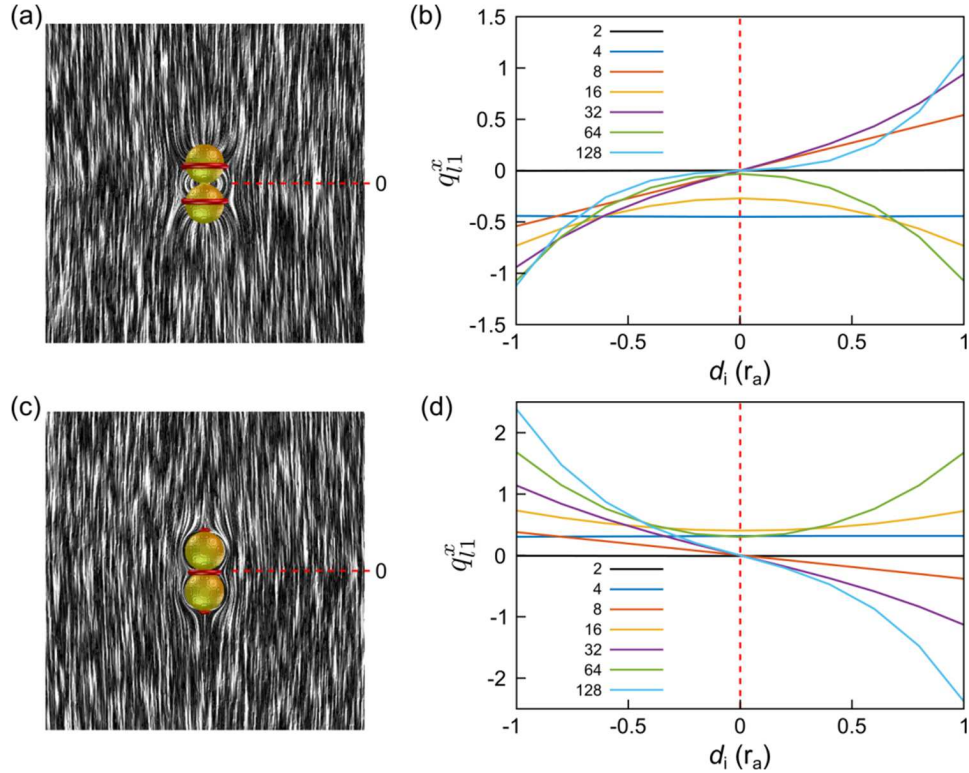

FIG. 10. Chain particle comprising of two equal spheres and its multipoles. (a) Chain colloid with homeotropic anchoring on the surfaces induces two Saturn rings. (b) The director field shows strong quadrupolar and hexadecapolar, and weak 64-polar contribution. Higher multipoles are zero in the geometrical center of composite chain colloid. (c) Chain colloid with tangential anchoring induces strong neck defect and two boojums at the poles. (d) The director field shows strong hexadecapole, quadrupole and 64-pole, whereas higher multipoles are zero in the geometrical center. The location of the geometrical center of the colloid is depicted with red dashed line (at $d_i=0$).

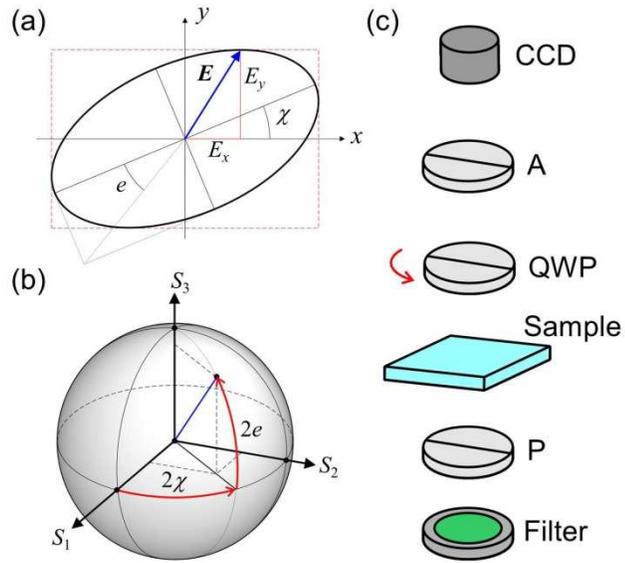

FIG. 11. Polarimetric imaging principles and setup. (a, b) Polarization ellipse and a Poincaré sphere for polarized light. (c) Experimental setup for measuring orientation and ellipticity of the polarization ellipse.